\documentclass[fleqn,usenatbib]{mnras}

\usepackage{newtxtext,newtxmath}

\usepackage[T1]{fontenc}

\DeclareRobustCommand{\VAN}[3]{#2}
\let\VANthebibliography\thebibliography
\def\thebibliography{\DeclareRobustCommand{\VAN}[3]{##3}\VANthebibliography}

\usepackage{graphicx}	
\usepackage{amsmath}	

\newcommand{\gizmo}{\textsc{gizmo}}
\newcommand{\slug}{\textsc{slug}}
\newcommand{\MSun} {\mbox{$M_{\odot}$}}

\newcommand{\aref}[1]{\hyperref[#1]{Appendix~\ref{#1}}}

\title[Actions turn awry: Diffusion in the Galactic Disc]{
{Fundamental limits to orbit reconstruction due to non-conservation of stellar actions in a Milky Way-like simulation}}
\author[A. Arunima et al.]{
Arunima Arunima,$^{1}$\thanks{E-mail: arunima.arunima@anu.edu.au}
Mark R. Krumholz,$^{1}$
Michael J. Ireland$^{1},$
Chuhan Zhang$^{1}$
and Zipeng Hu$^{2}$
\\
$^{1}$Research School of Astronomy and Astrophysics, Australian National University, Canberra ACT 2601, Australia \\
$^{2}$Kavli Institute for Astronomy and Astrophysics, Peking University, Beijing 100871, China
}

\date{Accepted XXX. Received YYY; in original form ZZZ}

\pubyear{\the\year{}}

\begin{document}
\label{firstpage}
\pagerange{\pageref{firstpage}--\pageref{lastpage}}
\maketitle

\begin{abstract}
The conservation of stellar actions is a fundamental assumption in orbit reconstruction studies in the Milky Way. However, the disc is highly dynamic, with time-dependent, non-axisymmetric features like transient spiral arms and giant molecular clouds (GMCs) driving local fluctuations in the gravitational potential on top of the near-axisymmetric background. Using high-resolution magnetohydrodynamic simulations that incorporate gas dynamics and star formation, we quantify the rate at which these effects drive non-conservation of the actions of young stars from Myr to Gyr timescales. We find that action evolution is well described as a logarithmic random walk, with vertical action evolving more rapidly than radial action; the diffusion rate associated with this random walk is weakly dependent on the stellar birth environment and scales approximately linearly with the galactic orbital frequency at a star's position. The diffusion rates we measure imply a fundamental limit of $\sim 100$ Myr as the timescale over which stellar orbits can be reliably reconstructed using methods that assume action conservation. By comparing diffusion rates for younger stars to those measured for an older and more vertically-extended control population, we conclude that radial action evolution is driven primarily by transient spiral arms, while vertical action evolution is driven by gravitational scattering off gaseous structures. Our results have significant implications for galactic archaeology and disc dynamics studies, necessitating a closer look at the timescales over which actions are assumed to be conserved in the disc.
\end{abstract}




\defcitealias{wibking2023}{WK23}
\defcitealias{chuhan2024}{Z25}
\defcitealias{hu2023}{H23}

\section{Introduction}
\label{sec:intro}

Understanding the dynamical evolution of stellar orbits is fundamental to reconstructing the past history of galaxies. The use of actions -- adiabatic invariants in an axisymmetric potential -- has been quite successful for this purpose in galactic studies. In the Milky Way, action-space analyses have been widely employed in studies identifying merger remnants and accreted substructures, as well as in globular cluster studies \citep{helmi2018,myeong2019,feuillet2020,lane2022, malhan2022,limberg2022,callingham2022,chen_gnedin2022,garcia2024}. While these approaches have been highly effective in the Galactic halo, where the environment is relatively dynamically quiet, applying action-based reconstruction methods to the disc presents additional challenges. 

The orbits of stars within the galactic disc are shaped by both external perturbations \citep{antoja2018,blandhawthorn2019,li2021,antoja2023,darragh-ford2023,frankel2023} and by secular internal processes such as interactions with giant molecular clouds (GMCs), spiral arms and the bar \citep{sellwoodbinney2002,roskar2012,vera-ciro2014,mackereth2019,tremaine2023}. These perturbations lead to deviations of the gravitational potential away from the time-invariant, axisymmetric state required for strict action conservation. Despite these complexities, efforts have been made to use actions for disc studies \citep{trick2019,coronado2020,coronado2022}. 

One prominent application of such methods is in cluster reconstruction studies that employ traceback techniques to infer the past positions and birth environments of stars. These methods typically assume a static, axisymmetric Galactic potential and rely on the conservation of stellar actions to extrapolate stellar trajectories backward in time. Numerous studies have used traceback methods to reconstruct the dispersal history of open clusters in the Milky Way disc, particularly in the Solar neighbourhood \citep{miret2018,miret2020,miret-roig22,galli2018,Crundall2019,Squicciarini2021,Heyl:2021b,Heyl:2021a,Schoettler:2020,Schoettler2021,Ma2022,zucker2022b,galli2023,couture2023,pelkonen2024}. 
However, the validity of these reconstruction studies depends critically on the assumption of individual stellar actions remaining conserved over the timescales involved.

Most existing studies of action evolution over time focus on the global distribution of actions, particularly in the context of radial migration and dynamical heating. Simulations have explored how secular evolution redistributes angular momentum and heats the disc over gigayear timescales \citep{roskar2008,vera-ciro2014,monari2016,vera-ciro2016,halle2018,mikkola2020,okalidis2022}. 
Observational studies similarly focus on population-wide properties, most notably through the age-velocity dispersion relation \citep[{AVR}, e.g.][]{sharma2014,yu2018,mackereth2019}, which is interpreted as evidence of cumulative dynamical heating from spiral structure, GMCs and other perturbations. A few observational studies have also studied action evolution by comparing the present-day actions of stars of different ages, again primarily in the context of disc heating and radial migration \citep{frankel2018,ting_rix_2019,frankel2020}.
However, these studies mainly characterise ensemble behaviour -- how the statistical distribution of actions or velocities broadens with time -- rather than the dynamical evolution of individual stellar actions. For applications such as orbit reconstruction, it is this individual time evolution of actions that is crucial. This is analogous to the distinction between the microscopic motion of gas molecules and the macroscopic evolution of temperature in a thermodynamic system -- such a system may have a completely constant, predictable temperature, yet it may still be impossible to backtrace the trajectory of any individual molecule over any substantial period. 

To quantify the time evolution of individual stars' actions, action diffusion provides a useful statistical framework. Prior studies have explored the long-term evolution of action due to scattering by spiral arms and bar-driven resonances \citep[e.g.,][]{solway2012,daniel2015,halle2018,kawata2021}. However, short-term diffusion, particularly in newly formed stars, remains largely unexplored. \citet{fujimoto2023} investigated GMC-driven scattering of stars on short timescales but did not compute actions, leaving open questions about perturbations in action space. Moreover, existing studies of action conservation have largely been conducted in $N$-body simulations, which lack gas dynamics and star formation \citep[e.g.,][]{solway2012,vera-ciro2016,mikkola2020}. Stars form in dense, turbulent gas clouds \citep{federrath2012,Krumholz2014}, where gravitational potential evolves rapidly due to gas accretion, stellar feedback and local dynamical instabilities. Hence, the initial conditions  of the stars are inherently linked to an evolving potential, making a self-consistent treatment of gas dynamics essential for studying early action evolution. Hydrodynamic plus $N$-body simulations that include self-gravity, radiative cooling and galactic scale gas flows, such as spiral arms and galactic shear, are the ideal tool for the purpose of exploring these effects.

These considerations motivate the present study, in which we use a high-resolution magnetohydrodynamics (MHD) simulation of a Milky Way-like disc galaxy to study the evolution of stellar actions. 
By computing these at high temporal resolution, we aim to measure the rate of action diffusion and characterise the timescales over which the actions deviate from conservation in the disc, providing insights into the reliability of traceback methods that rely on these assumptions. The remainder of this paper is organised as follows. In \autoref{sec:simulation}, we describe the galactic simulations used in this study, \autoref{sec:method} details our method used to calculate stellar actions and introduces key notation for studying their time evolution, \autoref{sec:results} presents our main results, including the distribution of actions, their temporal evolution, and dependence of the evolution on stars' birth environment, and in \autoref{sec:conclusion} we discuss the broader implications of our findings before summarising our conclusions. The second paper in this series will examine the use of actions to identify and reconstruct dissolved star clusters.

\section{Simulation}
\label{sec:simulation}

We analyse simulations of an isolated Milky Way-like disc galaxy with flocculent spiral structure and no bar taken from \citet[hereafter \citetalias{chuhan2024}]{chuhan2024}. This simulation is an extension of the full galaxy zoom-in simulations described by \citet{wibking2023} and \citet{hu2023} (hereafter \citetalias{wibking2023} and \citetalias{hu2023}, respectively). Here, we summarise the details of this simulation and direct readers to \citetalias{wibking2023}, \citetalias{hu2023} and \citetalias{chuhan2024} for further information. 

\subsection{Numerical method}
\label{subsec:num_methods}

The simulations solve the equations of ideal magnetohydrodynamics using the \gizmo~code \citep{hopkins2015,hopkins2016,hopkins_raives2016}, with gas and metal line cooling implemented via the GRACKLE library \citep{smith2017} (see Appendix A in \citetalias{wibking2023} for justification of this choice).
The treatment for star formation in the simulation is as follows: for gas particles with density $\rho_{\text{g}}$ exceeding a critical threshold $\rho_{\text{crit}}$, the local star formation density rate is calculated as 
\begin{equation}
    \dot{\rho}_{\text{SFR}} = \epsilon_{\text{ff}} \frac{\rho_{\text{g}}}{t_{\text{ff}}},
\end{equation}
where $\epsilon_{\text{ff}}$ is the star formation efficiency, $\rho_{\text{g}}$ is the local gas density and 
\begin{equation}
\label{eq:ff_time}
    t_{\text{ff}} = \sqrt{\frac{3\pi}{32G\rho}}
\end{equation} 
is the local gas free-fall time. We set $\epsilon_{\text{ff}}=0.01$, consistent with the mean value observed across a wide density range \citep{Krumholz_etal2019}. The critical threshold density $\rho_{\text{crit}}$ depends on the simulation resolution; as we discuss below, the simulation takes place in two stages, a low-resolution one to allow the galaxy to settle to statistical steady-state, followed by a higher resolution, shorter stage to capture more detail. During the initial phase $\rho_\mathrm{crit} = 100 \mu m_\mathrm{H}$ cm$^{-3}$, where $m_\mathrm{H}$ is the mass of a hydrogen atom and $\mu = 1.4$ is the mean mass per H nucleus for gas of standard cosmic composition; this value is chosen such that, for gas of density $\rho_\mathrm{crit}$ and at the equilibrium temperature implied by our cooling curve, the Jeans mass is nearly equal to the simulation mass resolution $\Delta M = 859.3$ M$_\odot$ -- see \citetalias{wibking2023} for details. During the second stage of the simulations, we increase $\rho_\mathrm{crit}$ to $1000\mu m_\mathrm{H}$ cm$^{-3}$, which roughly maintains this condition at the increased resolution. To avoid excessively small time steps and limit the computational expense in following very dense regions, we increase $\epsilon_{\text{ff}}$ to $10^6$ for gas particles with $\rho_{\text{g}} \geq 100 \rho_{\text{crit}}$, forcing them to be converted instantly into collisionless star particles. Star formation is implemented stochastically, such that the probability of a gas particle being converted to a star particle in a time step of size $\Delta t$ is
\begin{equation}
    P = 1 - \text{exp}(-\epsilon_{\text{ff}}\Delta t/t_{\text{ff}}).
\end{equation}
See \citetalias{chuhan2024} for further details.

Once star particles form, they interact with the galactic environment only through gravity and feedback. Since the resolution is high enough during the high-resolution simulation phase that each star particle does not represent an entire cluster sampling the full initial mass function (IMF), we cannot use the \gizmo~treatment of IMF-integrated stellar feedback. Instead, stellar feedback is determined on a star-by-star basis by forming each star particle from a number of individual stars, stochastically drawing a synthetic stellar population from a \cite{Chabrier2005} IMF using the \slug~stellar population synthesis code \citep{dasilva2012,krumholz2015}. The evolution of each star follows the Padova stellar tracks \citep{bressan2012}. The star's atmosphere is modelled using \slug's ``starburst99'' spectral synthesis method \citep{leitherer1999}. This provides us with each star's ionising luminosity as a function of its mass and age, which is injected back into the simulation using a Str\"omgren volume method. We also determine which stars end their lives as supernovae or asymptotic giant branch (AGB) stars, and inject feedback from these events following the recipe for handling partially-resolved SNe described by \citet{Hopkins:2018}. For full details on the treatment of the feedback see \citet{armillotta2019}. The mass and metal return in each time step for each star particle are based on its mass and evolutionary stage, following \cite{sukhbold2016} for type II supernovae, \cite{karakas_lugaro2016} for AGB stars, and \cite{doherty2014} for super-AGB stars. 

\subsection{Initial conditions}
\label{subsec:initial_cond}

As mentioned above, the simulation takes place in two stages. The first, described by \citetalias{wibking2023}, begins from the isolated Milky Way-analog AGORA project initial conditions \citep{Kim:2016} and run using an IMF-integrated treatment of feedback; 
\citetalias{chuhan2024} then use the 600 Myr snapshot from this simulation as the initial condition for their simulations. This snapshot exhibits a stable gas fraction similar to that of the present-day Milky Way. The \citetalias{wibking2023} simulations have gas particles with a mass of 859.3 M$_\odot$, dark matter particles with a mass of $1.254 \times 10^5$ M$_\odot$, and stellar disc and bulge particles with a mass of $3.4373 \times 10^3$ M$_\odot$. Star particles formed during the simulation inherit the mass of the gas particle from which they were created.
To enhance resolution from the original $\Delta M = 859.3$ M$_\odot$ to $286.4 M_\odot$, \citetalias{chuhan2024} first run without enhancing the resolution for 100 Myr, but using the star-by-star feedback prescription, to generate a realistic population of stellar particles for feedback and metal enrichment. They then increase the resolution using the particle splitting method described and used in \citetalias{hu2023}. As shown in Fig. 3 of \citetalias{chuhan2024}, the star formation rate initially spikes due to the resolution increase, but it then stabilises within approximately 100 Myr.

\begin{figure}
\centering
        \includegraphics[width=0.48\textwidth]{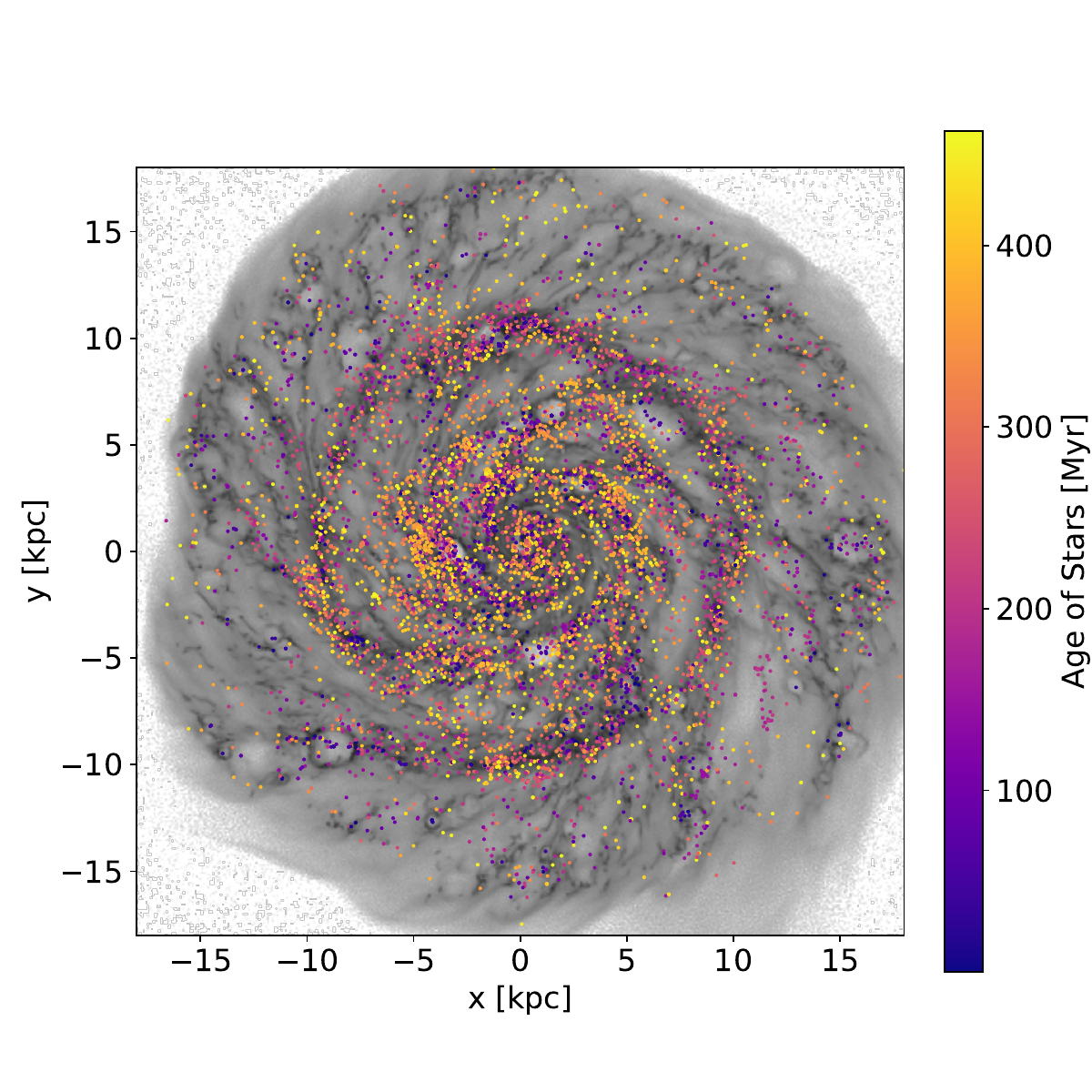}
        \caption{A snapshot of the simulation at $t=464$ Myr (the final snapshot). We show the log of the gas surface density in greyscale in the background, with darker shades corresponding to higher densities. Overlaid on this is a random subsample representing 1\% of the total star sample we use for our analysis, with the colour of each particle indicating its age as shown by the accompanying colour bar.}
        \label{fig:snapshot}
\end{figure}

We begin the analysis we describe below at this point, which for the purposes of this paper we define as $t=0$. The simulations run for 464 Myr from that point, with output snapshots written at 1 Myr intervals, providing us with a data set of this duration and $\approx 300$ M$_\odot$ mass resolution to study the star particle dynamics. At $t=464$ Myr, we have approximately 1.32 million star particles. In \autoref{fig:snapshot} we show 1\% of the total star sample, selected at random. The greyscale background shows the gas surface density, with darker regions indicating higher densities. The overlaid stars are colour-coded according to their ages, as indicated by the accompanying colour bar, showcasing the spatial distribution and age variation of the stellar population.
It is important to note that these ``star particles'' do not represent individual stars. Simulations resolving down to individual stars that sample the entire IMF are not feasible at present. However, they are small enough aggregates of stars that we can analyse them to examine the dynamical information that they retain about their birth properties as if they were individual stars.

\section{Calculating stellar actions}
\label{sec:method}
This section outlines how we calculate stellar actions from simulation outputs and study their evolution. A basic outline of our procedure is that we use the potential output for each particle to build an axisymmetric model for the galaxy's gravitational potential, and then use the time-dependent position and velocity of each star particle to compute its actions. The following subsections provide full details of the method.

\subsection{Gravitational potential profile}
\label{subsec: grav pot profile}
To accurately model the time-dependent gravitational potential of the galaxy, we first define an appropriate reference frame. We cannot simply use the simulation frame because, although the simulation is initialised with the galactic plane at rest at $z=0$ and cylindrically symmetric about the origin, over the duration of the simulation supernova explosions drive an asymmetric wind off the galaxy, imparting a non-negligible momentum to the remaining gas and to the stars to which it is gravitationally bound.

\begin{figure}
        \includegraphics[width=0.47\textwidth]{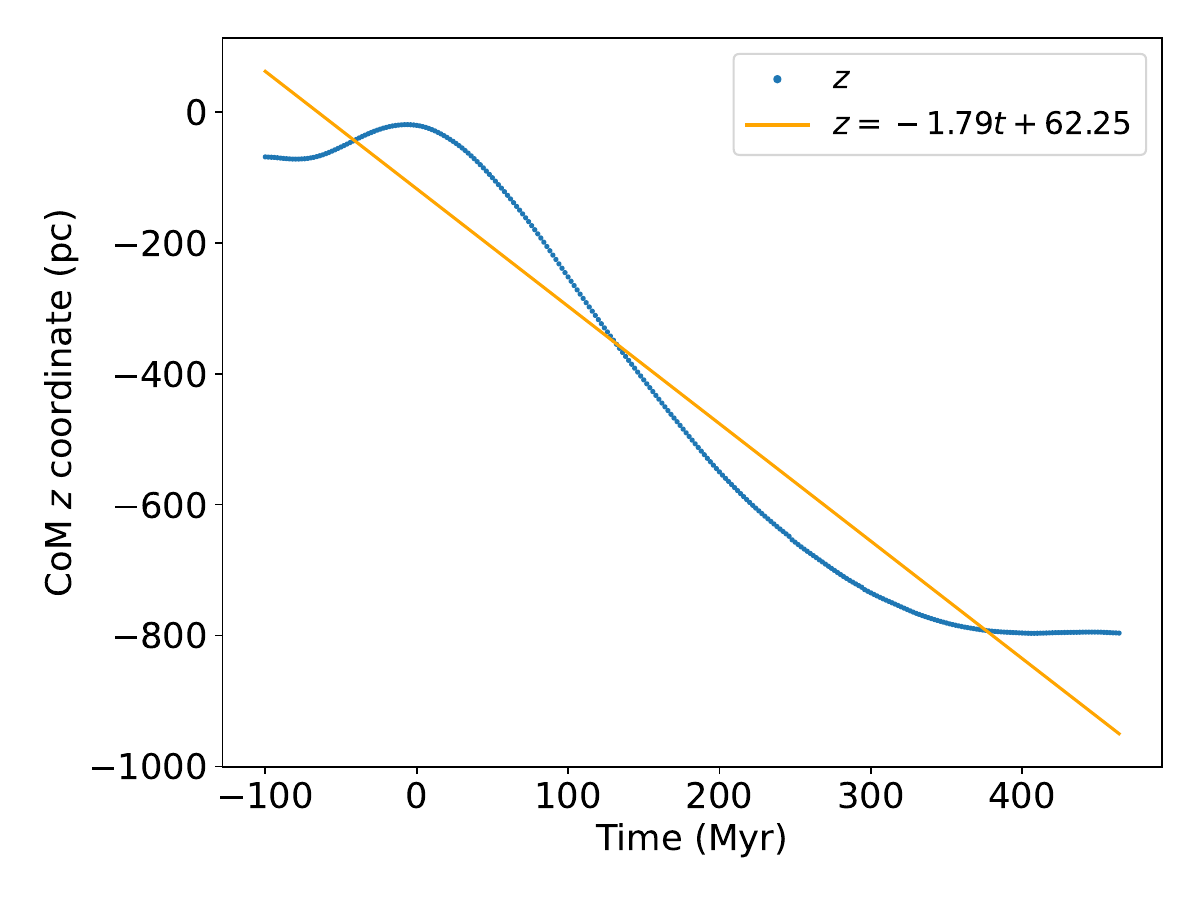}
        \caption{Coordinates of the centre of mass of initial stars in the simulation with time in the $z$ direction. We also show a least-squares linear fit to the $z$ position (solid line), the functional form for which is provided in the legend (with position in units of pc and times in units of Myr); the $R^2$ value for the fit is 0.94.}
        \label{fig:com}
\end{figure}

To correct for this shift, we calculate our actions with respect to the rest frame of the galaxy. To this end, we use the population of stellar disc particles that are already present in the initial conditions of \citetalias{chuhan2024} (as described in \autoref{subsec:initial_cond}). We refer to these pre-existing particles as `initial stars'. Having calculated the centre of mass (COM) of the initial stars at each snapshot, we plot the resulting position in $z$-direction as a function of time in \autoref{fig:com}.

We observe a velocity $\sim 2$ km/s in the $z$ direction and $\sim 1$ km/s in the $x$ and $y$ directions, and this movement of the COM of the initial stars is consistent with what we expect: the simulation produces galactic winds with a mass flux $\sim 1$ $\MSun$/yr and a velocity of a few hundred km/s (\citetalias{wibking2023}), so over the simulation timescale of $\sim 500$ Myr winds eject $\sim 5 \times 10^{8} \MSun$, roughly 2\% of the total initial stars' mass $\approx 3 \times 10^{10} \MSun$. This results in a net velocity of the stellar and gas disc of order of a few km/s. The $z$ velocity is greater than the $x$ and $y$ velocities due to the greater escape of winds normal to the galactic plane.


The combination of lower velocity in the $xy$ plane and the much larger extent of the disc in this direction mean that the displacements in the plane can be safely ignored. By contrast, this is not true in the $z$ direction, with the displacement of $\approx 0.8$ kpc over the course of the simulation is several times larger than the scale height of the thin disc. To remove this drift, we carry out a linear fit to the $z$ position as a function of time, which we show as the straight orange line in \autoref{fig:com}. This line defines a time-dependent COM position and constant velocity for the galaxy relative to the simulation frame, and our first step in computing the galactic potential is therefore to shift all particle positions and velocities for all time snapshots into this frame. However, we note that the linear fit is clearly not a perfect representation of the actual COM position as a function of time, which is not surprising, since acceleration of the galaxy by recoil from galactic winds applies a stochastically-varying acceleration. This potentially complicates the situation, as it suggests the absence of an inertial frame in which the galactic plane remains at rest. Such a scenario foreshadows the non-conservation of stellar actions, which would be a fundamental issue since most analyses rely on the assumption of a stable inertial frame for dynamical calculations \citep[e.g.,][]{bovy2015,sanders_binney2016}. We return to this point below.

\begin{figure}
\centering
        \includegraphics[width=0.48\textwidth]{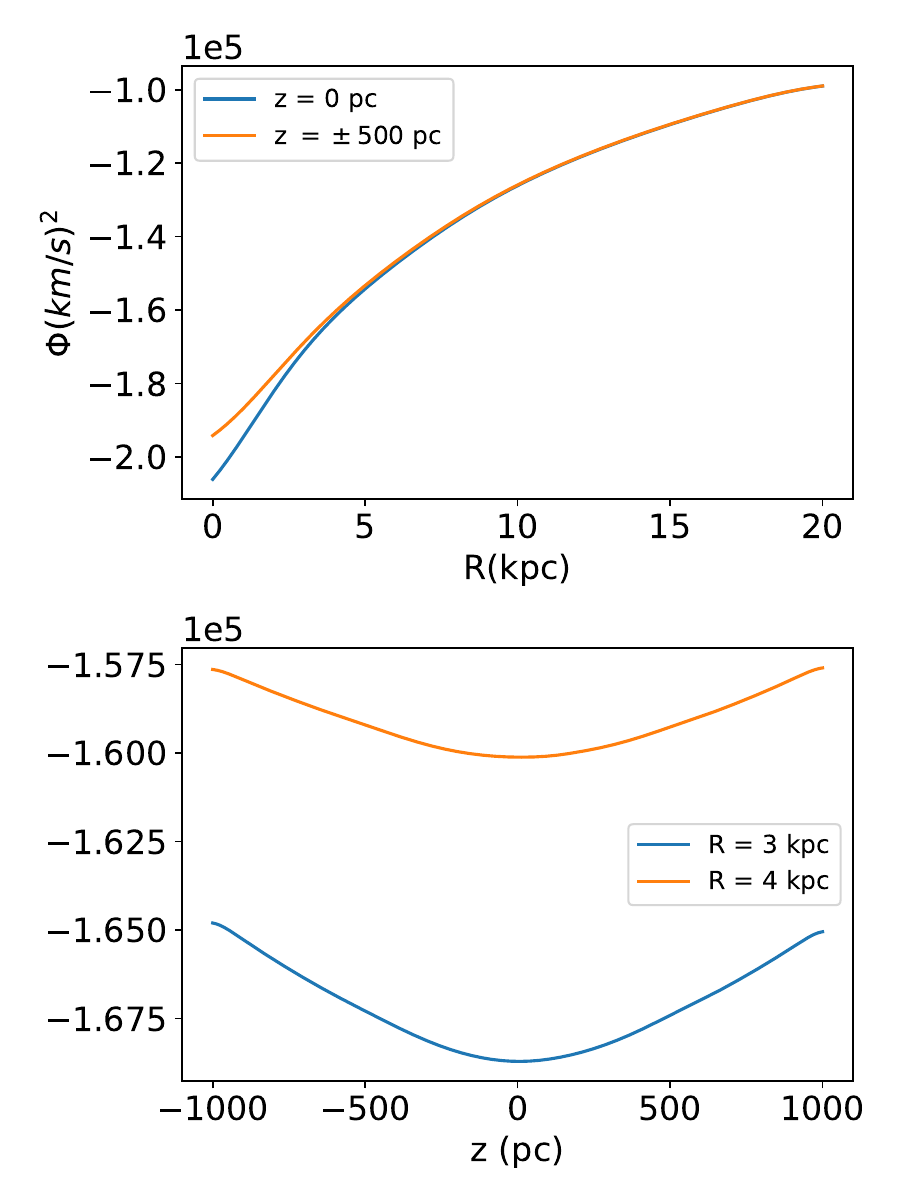}
        \caption{Variation in potential $\Phi$ with $R$  at sample $z$-values (top), and with $z$ at sample $R$ values (bottom), for the $t=100$ Myr simulation snapshot.}
        \label{fig:potential}
\end{figure}

Having established our reference frame, we are now ready to compute our best estimate of an azimuthally symmetric galactic potential from large but sparse samples at our star particle positions. To do so, for each simulation snapshot we define a grid in cylindrical coordinates that spans the entire galaxy, with radial coordinates extending from 0.1 pc to 20 kpc, a vertical range of $\pm 1$ kpc, and a resolution of 1 pc in both directions. The grid in the azimuthal direction has a resolution of $2\pi/25$ radians. For each point on the defined grid, we identify the nearest-neighbour particle and assign its potential value (which is computed by the \textsc{gizmo} gravity solver) to that grid point. We then average over the $\phi$ direction to create an axisymmetric 2D potential grid in $(R,z)$. \autoref{fig:potential} shows some sample slices through the 2D potential we derive for one time snapshot.

\subsection{Action calculation}
\label{subsec: action calc}

Our next step is to calculate the actions of each star. Performing this calculation in full generality requires computationally-expensive numerical integration, an approach that is impractical given the size of our data set, $\sim 10^8$ stellar particles. However, we are solely concerned with stars that are both young -- age $\lesssim 0.5$ Gyr -- and near the galactic plane -- root mean square height $\langle z^2\rangle^{1/2} <200$ pc, and $|z| < 1$ kpc for all stars. For stars with these properties, the epicyclic approximation, whereby we decompose stellar orbits into independent radial and vertical oscillations about a guiding center, is highly accurate \citep{hunt2025}. Quantitatively, \citet{solway2012} find that, even in the presence of spiral structures, for stars up to 1 kpc off the plane, and over a timescale of a few Gyr, vertical actions computed using the epicyclic approximation change by 20.7\%, compared to 15.6\% with a more exact calculation; this small difference confirms that the epicyclic approximation is reliable for stars near the plane, particularly over the timescales of interest in our study.
To further validate the suitability of this approximation for our dataset, in \aref{app:galpy} we compare actions for a representative sample of stars computed using the epicyclic approximation to results we obtain using a more precise, but computationally intensive, orbit integration method. This comparison confirms that the errors introduced by the epicyclic approximation are small compared to the signals we measure.

\begin{figure*}
        \includegraphics[width=\textwidth]{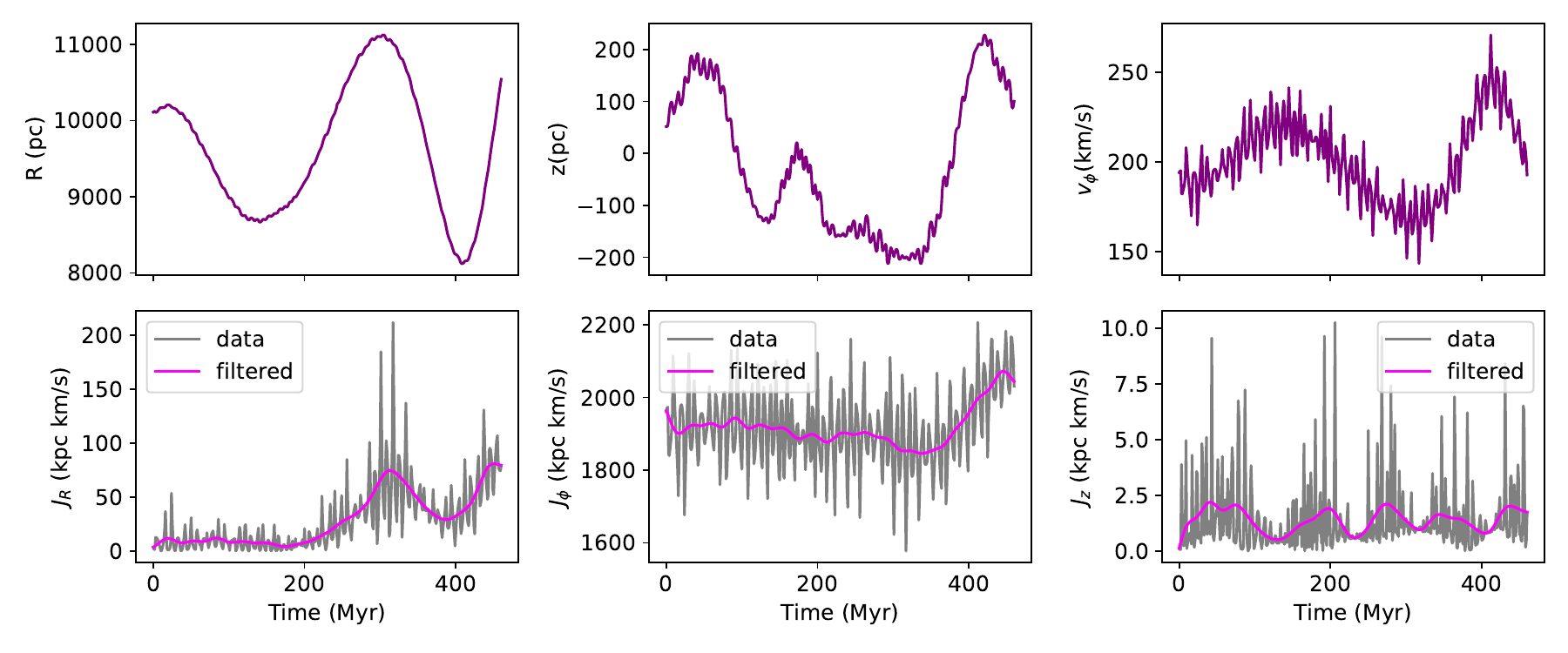}
        \caption{An example of a star that exhibits rapid variations in actions, likely because it is part of a bound star cluster. The top three panels show the $R$ and $z$ position  and $v_{\phi}$ velocity as a function of time, while the bottom three show actions versus time. The grey lines represent the actual actions we compute, while the pink lines show the actions after applying the low-pass filter as discussed in \autoref{subsec: action calc}.}
        \label{fig:orbit}
\end{figure*}
Given these considerations, we proceed using the expressions for the epicyclic approximation provided by \citet{binney2008}. In this approximation, stellar oscillations are characterised by the radial ($\kappa$) and vertical ($\nu$) epicyclic frequencies, while the azimuthal rotation about the galactic centre is defined by the $z$-component of angular momentum $L_z$. The guiding centre radius $R_g$ is related to the angular momentum $L_z$ as
\begin{equation}
 |L_z| = R_g^2 \Omega (R_g)
 \label{eq:Rg}
\end{equation} where 
\begin{equation}
    \Omega (R) = \sqrt{\frac{1}{R} \left. \frac{\partial \Phi}{\partial R} \right|_{R, z=0}}
    \label{eq:Omega}
\end{equation}
is the circular frequency. The epicyclic frequencies are related to the potential as
\begin{eqnarray}
    \kappa^2 & = & \left( R \frac{\partial \Omega^2}{\partial R} + 4\Omega^2\right)_{(R=R_g, z=0)} 
    \label{eq:kappa} \\
    \nu^2 & = & \left. \frac{\partial^2 \Phi}{\partial z^2}\right|_{(R=R_g, z=0)}.
    \label{eq:nu}
\end{eqnarray}
We compute $R_g$, $\kappa$ and $\nu$ for each star as follows. We first use the star's radial position $R$ and azimuthal velocity $v_\phi$ to evaluate its angular momentum $L_z = R v_\phi$, and then to use \autoref{eq:Rg} together with the galactic rotation curve $\Omega(R)$ to evaluate the star's guiding radius $R_g$. Once $R_g$ is known, we can then evaluate $\kappa(R_g)$ and $\nu(R_g)$ from \autoref{eq:kappa} and \autoref{eq:nu} using the known rotation curve and potential. The primary challenge to executing this strategy is that the gravitational potential from the simulation is discrete and subject to numerical noise, and thus considerable care is required when evaluating the derivatives that appear in the expressions above. We provide a detailed description of our full methodology in \aref{app:action_calc}.


Finally, we find expressions for radial, vertical, and azimuthal actions in the epicyclic approximation. In an axisymmetric potential, the azimuthal action $J_{\phi}$ is defined as
\begin{equation}
    J_{\phi} = \frac{1}{2\pi} \oint v_{\phi} R \,d\phi .
\end{equation}
Using $v_{\phi} = R\dot{\phi}$ and $L_z = R^2\dot{\phi}$, and solving the above integral, we get
\begin{equation}
    J_{\phi} = L_z.
\end{equation}
For the purpose of computing the radial and vertical actions, we note that in the epicyclic approximation we treat the radial and vertical motion as two independent harmonic oscillators, so that the Hamiltonian for the perturbation about the circular orbit is
\begin{equation}
    H_\mathrm{pert} = \frac{v_R^2}{2} + \frac{\kappa^2}{2}(R-R_g)^2 + \frac{v_z^2}{2} + \frac{\nu^2}{2}z^2
\end{equation}
where $v_R$ and $v_z$ are the velocities in the radial and vertical directions, respectively, and $z$ is the vertical coordinate of the star. The potential energy terms $(\kappa^2/2)(R-R_g)^2$ and $(\nu^2/2)z^2$ correspond to the radial and vertical oscillations around the equilibrium orbit. The usual simple harmonic oscillator has Hamiltonian $H = \frac{1}{2}(v^2 + \omega^2 x^2)$ for velocity $v$, displacement $x$, and oscillator frequency $\omega$, and the corresponding action is $E/\omega$, where $E$ is the energy. By analogy, for our epicyclic approximation Hamiltonian the radial energy $E_R$ is
\begin{equation}
    E_R = \frac{v_R^2}{2} + \frac{\kappa^2}{2}(R-R_g)^2.
\end{equation}
and the corresponding radial action is
\begin{equation}
    J_R = \frac{E_R}{\kappa} = \frac{v_R^2 + \kappa^2 ( R - R_g)^2}{2\kappa},
\end{equation}
Similarly, the vertical energy $E_z$ is
\begin{equation}
    E_z = \frac{v_z^2}{2} + \frac{\nu^2}{2}z^2,
\end{equation}
and the vertical action $J_z$ is 
\begin{equation}
   J_z = \frac{v_z^2 + \nu^2z^2}{2\nu}.
\end{equation}
Thus given the radial and vertical epicyclic frequencies and guiding radii computed above, together with the stellar radial and vertical velocities, we can compute the radial and vertical actions for all stars.



\begin{figure*}
        \includegraphics[width=\textwidth]{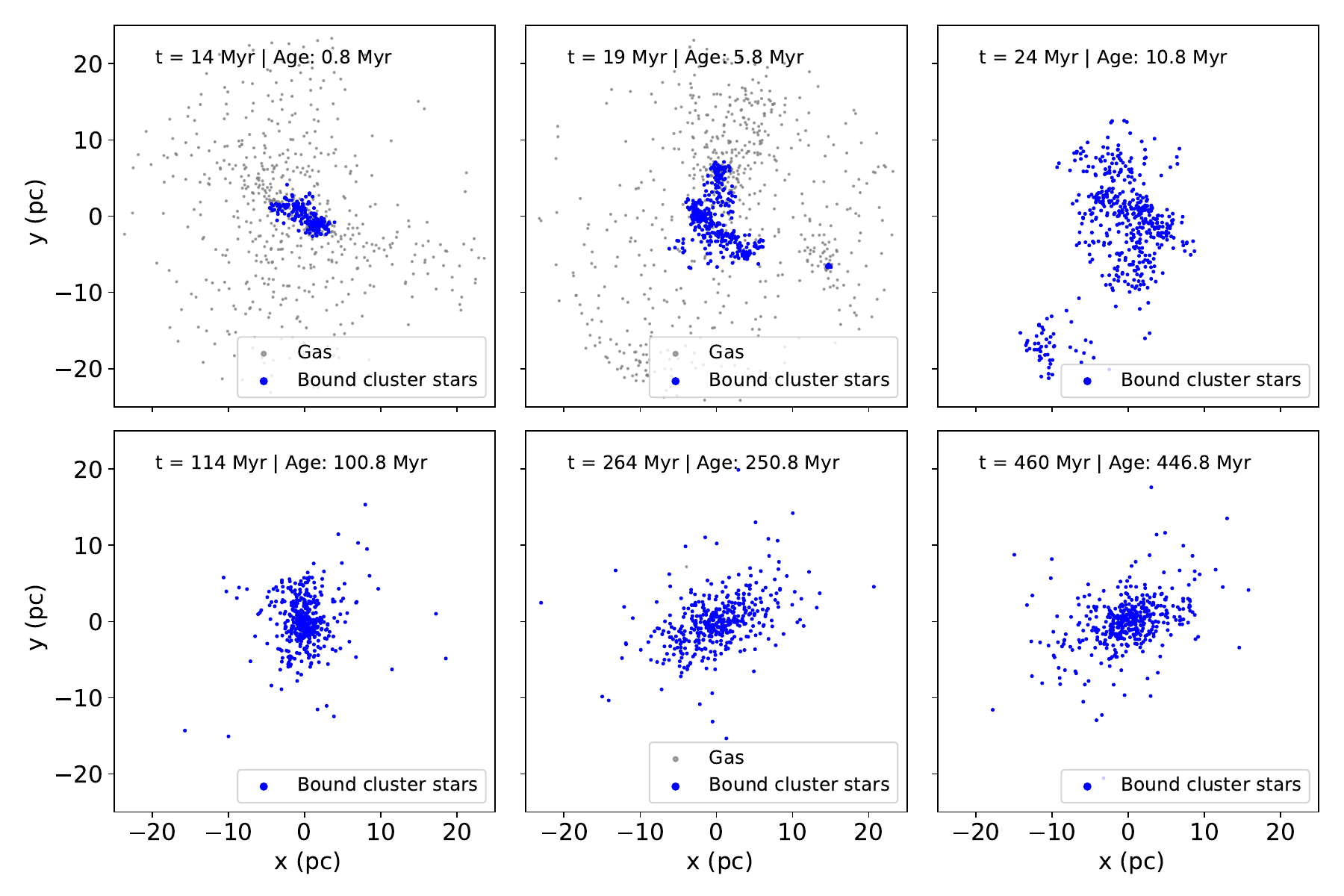}
        \caption{An example of a bound cluster in our simulation. We plot the $x$ and $y$ coordinates of the member stars (blue points) relative to the cluster's centre of mass, showing their spatial configuration in six different snapshots. The snapshot time $t$ and the mean stellar age of the stars shown is annotated in each panel. Grey points show gas particles.}
        \label{fig:cluster}
\end{figure*}

Once we have computed the actions for all stars at each snapshot, we are in a position to investigate their temporal evolution. Initial analysis reveals that a significant fraction of stars exhibit rapid variations in their actions over short  (few Myr or less) timescales.
These fluctuations are not caused by large changes in radial or vertical position but by substantial variations in azimuthal velocity $v_{\phi}$, as shown in \autoref{fig:orbit}, which occur because the stars that exhibit this behaviour are part of gravitationally bound star clusters. We illustrate an example of such a bound structure in \autoref{fig:cluster}. We identify cluster members as follows: we start with the example star that exhibits rapid action variations (in this case the one shown in \autoref{fig:orbit}), and then we iteratively add the next-nearest star and check whether the resulting system is gravitationally bound by computing the total energy (kinetic + potential energy) of the system. We continue adding stars in increasing order of their distance from the original star until the inclusion of a new member renders the total energy positive. We do this at the last snapshot. We then track the resulting set of bound members backwards in time, and re-evaluate their boundedness at each of the snapshots shown in \autoref{fig:cluster}. Thus the stars shown in each panel consist of those that are bound to our example star both at the time shown and at the final snapshot, and the persistence of large numbers of members between times indicates that we are examining a gravitationally bound population that remains together for many internal dynamical times. All of the stars we have examined that exhibit rapid action variations prove to be parts of structures such as this. In these gravitationally bound clusters, internal orbital motion about the cluster's centre of mass drives coherent velocity changes in the galactocentric frame that we see in \autoref{fig:orbit}. These internal dynamics, while physically real, are not indicative of long-term secular evolution in the galactic potential which is the primary focus of this study. Such bound structures will be the principal focus of Paper II in this series. For the moment, to isolate the long-term behaviour from the high-frequency variations, we apply a Butterworth filter with a cut-off frequency of 1/30 Myr$^{-1}$ to the time series of actions, thereby suppressing variations occurring on timescales shorter than 30 Myr. As shown in \autoref{fig:orbit}, this effectively smooths out the short-timescale oscillations caused by internal cluster dynamics while preserving secular trends in actions.


\section{Simulation Validation}
\label{sec:comparison}
Before we examine how the actions of individual stars evolve -- the primary purpose of this paper -- we first validate our simulation and our method for calculating actions by comparing to past studies that have focused on the evolutions of actions or closely related quantities for stellar populations rather than individual stars. We begin in \autoref{subsec:J_dist} by examining how the distribution of orbital actions for stars of different ages in our simulation compare to observations. We then perform a similar check in \autoref{subsec:avr} by comparing the trend of stellar velocity dispersion (which is coarsely related to the action distribution) with age in our simulations to previous results.

\begin{figure*}
\centering
        \includegraphics[width=0.9\textwidth]{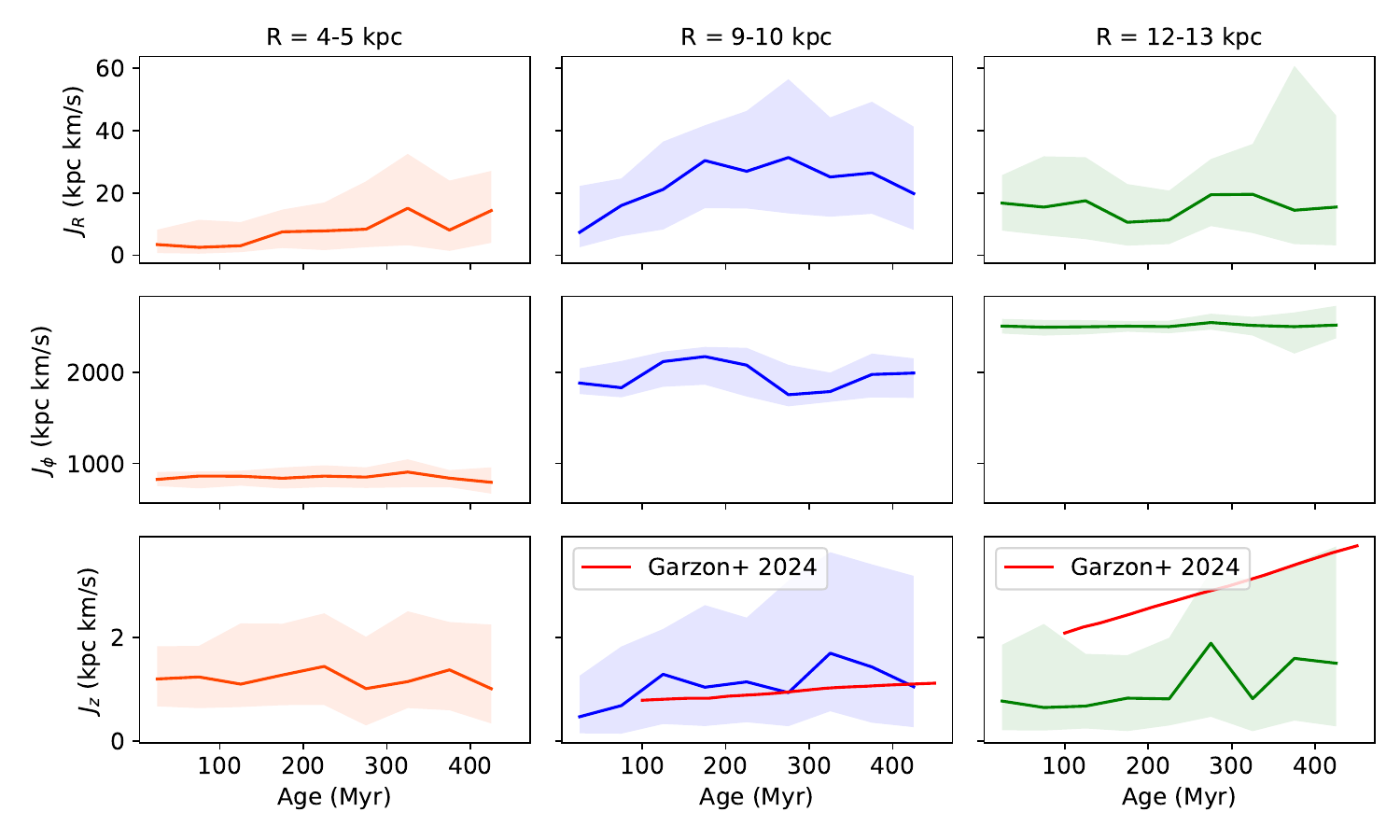}
        \caption{Distribution of actions (row-wise in order: $J_R$, $J_{\phi}$ and $J_z$) for stars of different ages at the last snapshot, $t=464$ Myr, in the radial bins 4-5 kpc (left), 9-10 kpc (middle), and 12-13 kpc (right). The orange, blue and green lines represent the medians of the distributions, with shading indicating the 16th and 84th percentile range. The red lines in the two right-most columns of the bottom row show the mean $J_z$ values as a function of age reported by \citet[their Figure 6]{garzon2024}.}
        \label{fig:dist_J}
\end{figure*}

\subsection{Distribution of actions}
\label{subsec:J_dist}

In order to compare the action evolution in our simulations to observations, we select our final snapshot ($t=464$ Myr) and bin the stars in that snapshot by radial position $R$ and age $\tau$ (defined as $\tau = t-t_\mathrm{form}$, where $t_\mathrm{form}$ is the time at which each star formed); this provides a snapshot of the data that can be compared directly to observations. In \autoref{fig:dist_J} we show the distributions of radial ($J_R$), azimuthal ($J_{\phi}$), and vertical ($J_z$) actions as a function of $\tau$ for the 4-5 kpc, 9-10 kpc, and 12-13 kpc radial bins. The solid central lines represent the medians of these distributions, while the shaded areas show the 16th to 84th percentile range. 

Our results reveal that the overall distribution of actions remains relatively stable over the simulation period. We find that $J_\phi$ is essentially static, which is not surprising, given that our $\approx 500$ Myr run time is too short for significant radial mixing. There is some weak evolution of $J_R$ and $J_z$ distributions, and the latter is particularly useful because there have been several observational studies of this evolution to which we can compare. \citet{ting_rix_2019} examine this quantity for stars of age up to 8 Gyr and observe significant broadening of the distribution, but this timescale is much larger than that for which our simulations run and is therefore not directly comparable. More recently,  \citet{garzon2024} performed a similar analysis for stars younger than 0.5 Gyr, well-matched to the timescale of our simulation. We overlay their results in the bottom row of our plot (in red). Since their analysis is limited to stars with  Galactocentric radii from 8 to 13 kpc, we only have comparisons for two of the radial bins plotted. For the 9-10 kpc radial bin, we find that our results agree very well with their mean $J_z$ values across ages. However, in the 12-13 kpc radial bin, we find significantly lower actions and less evolution. \citet{garzon2024} suggest that the very large change in action evolution from $9-10$ kpc to $12-13$ kpc that they find may be due to the warping of the outer Milky Way disc \citep[e.g.,][and references therein]{uppal2024}, a feature that is absent from our simulations, since we begin with a flat disc and do not include perturbations from dwarf galaxies or dark matter sub-halos that might induce a warp. This is likely the reason that we do not reproduce this aspect of the observations in our simulations. This also means that the results we derive below for the rates at which individual stellar actions evolve are likely to be underestimates in the far outer disc beyond $\sim 10$ kpc where the warp is significant. We should therefore think of the results we obtain as lower limits that describe action evolution in the absence of large-scale external perturbations to the galactic potential.

\begin{figure}
        \includegraphics[width=0.47\textwidth]{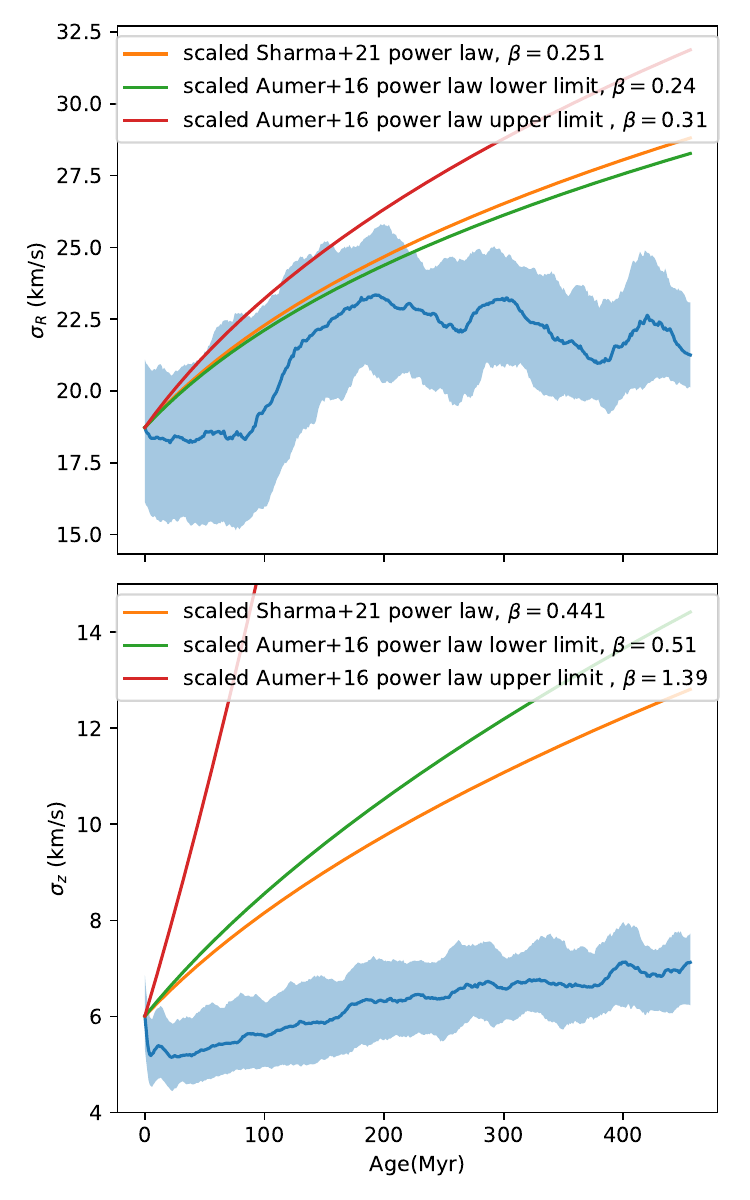}
        \caption{{Stellar velocity dispersion versus age (top: radial velocity dispersion; bottom: vertical velocity dispersion). The blue line and band show the median and 16th to 84th percentile range from our simulations, computed as described in the main text. The orange line shows the power law relation from \citet{sharma2021}, rescaled and extrapolated to the younger ages in our simulation, while the green and red lines show the lower and upper limits, respectively, of the scaled power law relations found by \citet{aumer2016} in their simulations. The $\beta$ values for each line are given in the legend.} }
        \label{fig:vel_disp}
\end{figure}

\subsection{Age-velocity dispersion relation}
\label{subsec:avr}

Stellar velocity dispersions are closely related to the distributions of stellar actions, and thus the age-velocity dispersion relation (AVR) in our simulations also offers the opportunity for a (slightly less direct) test of action evolution. We therefore compare the AVR in our simulations to two previous studies: the observational results of \citet{sharma2021} and the $N-$body simulations of \citet{aumer2016}. However, we caution that, unlike in our analysis of the action distribution from \citet{garzon2024} in \autoref{subsec:J_dist}, this comparison can only be rough and qualitative, because the data sets available are much less well-matched to our simulations. \citeauthor{sharma2021}'s observational study contains no measurements for stars younger than 2 Gyr, and so to compare to our study over $\approx 0.5$ Gyr we must extrapolate their trends well beyond the range of the data. Similarly, \citeauthor{aumer2016} do not include star formation and stellar feedback, and include gas only in a subset of their simulations and only then with a fixed temperature of $10^4$ K rather than realistic heating and cooling; they rely on a subgrid model to attempt to capture the effects of the omitted physics. We should therefore not necessarily expect good agreement between their results and ours.

With these caveats understood, we start by computing the AVR in our simulation as follows: we first filter the stellar velocities using the same Butterworth filter we apply to actions (\autoref{subsec: action calc}) to remove short-timescale oscillations associated with virial motion in bound clusters. Next, for each simulation snapshot, we compute the age $\tau$ of each star in that snapshot, and bin the stars by age in 1 Myr-wide bins. Within each bin, we compute the radial and vertical velocity dispersion, yielding values of $\sigma_{R,z}$ versus $\tau$ for that snapshot. Repeating this procedure for every snapshot yields multiple measurements of $\sigma_{R,z}(\tau)$, with the number of measurements decreasing with $\tau$ -- for $\tau = 1$ Myr we have 464 measurements (since each output snapshot includes stars that are $0-1$ Myr old), while for $\tau = 464$ Myr we have only a single measurement (since only the final output contains stars $463-464$ Myr old). From this set of measurements, we compute the 16th, 50th and 84th percentiles of the resulting $\sigma_{R,z}$ distributions, which we plot as a function of $\tau$ in \autoref{fig:vel_disp}. 

For the purposes of comparing to the results from \citet{aumer2016} and \citet{sharma2021}, we note that these authors fit the AVR to an empirical power law functional form 
\begin{equation}
    \label{eq:power_law}
    \sigma_{R,z} (\tau) = C_{R,z} \left (\frac{\tau + 0.1\,\mathrm{Gyr}}{10.1\,\mathrm{Gyr}} \right)^{\beta_{R,z}}
\end{equation}
where $C_{R,z}$ is a scaling constant and $\beta_{R,z}$ is the corresponding power-law exponent. The scaling constant is not particularly meaningful, since for \citeauthor{aumer2016} this is set primarily by their subgrid model (in which the initial stellar velocity dispersion is a free parameter), while for \citeauthor{sharma2021} our simulations are well outside the range of ages they have measured. For this reason, we rescale the values of $C_{R,z}$ found by these authors to match our results at $\tau = 0$, allowing us to focus on the trend with time rather than the absolute values. We plot the resulting rescaled curves in \autoref{fig:vel_disp}; for \citeauthor{aumer2016}, we plot two curves, corresponding to the minimum and maximum values of $\beta$ they determine by varying the parameters of their subgrid models.

We find that the velocity dispersions of our simulation grow more slowly with age than in either comparison studies. For $\sigma_z$ the measured result from \citeauthor{sharma2021} is intermediate between the results of our simulations, which predict slower $\sigma_z$ growth, and the N-body simulations of \citeauthor{aumer2016}, who predict faster-than-measured growth. The slower growth of the AVR in our simulations compared to the observational results from \citet{sharma2021} may reflect the fact that their fits span $\approx 10$ Gyr, over which time the Milky Way was likely significantly more dynamically active -- due to higher gas content and merger activity -- than in the present-day state that we simulate. Our differences from \citeauthor{aumer2016} almost certainly arise due to the differences between our self-consistent treatment of gas and stellar feedback versus their subgrid model.

The most important point to take away from \autoref{fig:vel_disp}, however, is that our simulations are quite dynamically quiet, with slow growth of the AVR. Significant changes appear to require Gyr timescales, here as well as in the previous studies. In \autoref{sec:results}, we contrast this slow, population-averaged evolution with the evolution of the individual stellar actions, which we show occurs on a much shorter timescale.

\section{Results}
\label{sec:results}
For characterising the time-evolution of stellar actions, we first define some statistical measures of the same at the beginning of \autoref{subsec:del_J}.
We then study the properties of the time series describing individual stars rather than the full population. We discuss how the evolution varies with the environmental conditions under which stars are born in \autoref{subsec:density} and \autoref{subsec:radius}. In \autoref{subsec:initial_stars_comparison}, we compare the action evolution for two different population of stars.

\subsection{Change in actions over time}
\label{subsec:del_J}

To characterise the change in action over time, we compute the absolute and relative differences in each component of the action between every possible pair of snapshots. Formally, for each star we define the absolute difference as
\begin{equation}
    \Delta J_i (t,\Delta t) = \left | J_i(t+\Delta t) - J_i(t) \right |
    \label{eq:DeltaJAbs}
\end{equation}
where $i = (R,z,\phi)$ is the component, $t$ is the time of the snapshot, $\Delta t$ is the time interval between the two snapshots and $J_i(t)$ is the $i$ component of the star's action action at time $t$ in the simulation. Similarly, we define the relative difference\footnote{Astute readers might worry that computing relative difference in action could yield infinite values for stars with very small initial actions (e.g., those born on nearly circular orbits with $J_R \simeq 0$).  While this is true in principle, our analysis avoids this problem by using median statistics, which are robust to such outliers and hence, we find that the relative changes remain well-behaved.} as
\begin{equation}
    \delta J_i (t,\Delta t) = \frac{\Delta J_i (t,\Delta t)}{J_i(t)}.
    \label{eq:DeltaJrel}
\end{equation}
We have $\approx 450$ time snapshots, and therefore $\approx 10^5$ snapshot pairs, and $\sim 10^6$ stars, so our total sample for analysis consists of $\sim 10^{11}$ action differences. While this gives us a very large sample size for statistical robustness, it is important to note that these measurements are not all independent as they include correlated measurements of the same stars at different times. 

We are interested in how stellar actions change with time difference $\Delta t$, but also with stellar age. To facilitate this, we note that each star has a formation time $t_\mathrm{form}$, and for each action pair $(t,\Delta t)$ we define $t_* = t - t_\mathrm{form}$ as the age of the star at the \textit{earlier} time -- thus for example if we are computing differences in action between snapshots at 100 Myr and 250 Myr, and we are examining a star that formed 95 Myr into the simulation, then we have $t = 100$ Myr, $\Delta t = 150$ Myr, and $t_* = 5$ Myr. Thus our action differences $\Delta J_i$ or $\delta J_i$ have both a time gap $\Delta t$ and an initial stellar age $t_*$ associated with them, and below we will bin our data by $\Delta t$ and $t_*$.

\begin{figure}
        \includegraphics[width=0.47\textwidth]{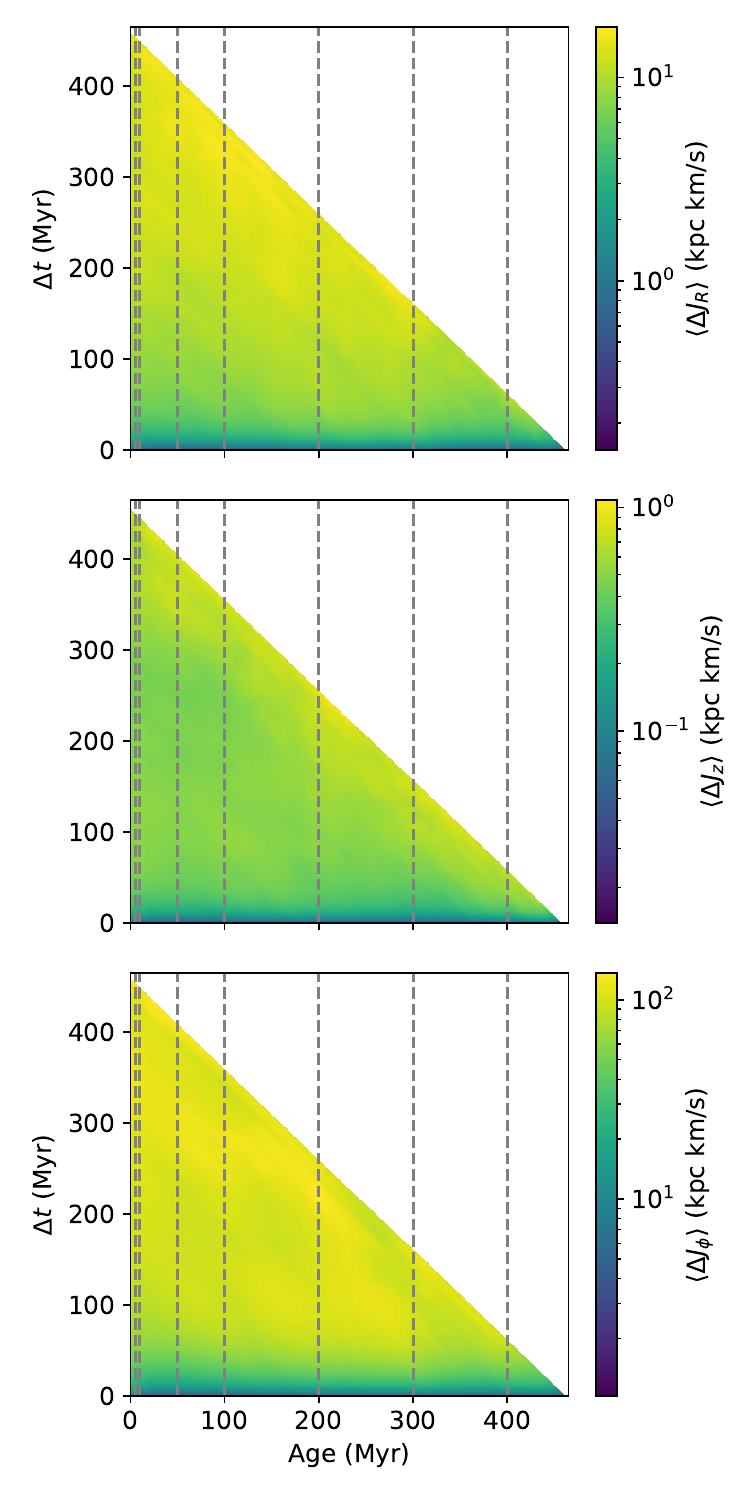}
        \caption{Heatmap of the median of the absolute change in stellar actions (row-wise in order: $\Delta J_R$, $\Delta J_z$, and $\Delta J_{\phi}$) as a function of the time lag ($\Delta t$) and initial stellar age ($t_*$). The dotted lines indicate initial stellar ages of 5, 10, 50, 100, 200, 300 and 400 Myr, corresponding to the time slices plotted in \autoref{fig:abs_J}. The empty triangle in the top right of the plot corresponds to combinations of initial stellar age and time lag that are inaccessible due to the 464 Myr duration of the simulation that we analyse.} 
        \label{fig:heatmap}
\end{figure}

We now investigate the evolution of individual stellar actions over time as characterised by the absolute changes $\Delta J_R$, $\Delta J_z$ and $\Delta J_{\phi}$ (\autoref{eq:DeltaJAbs}). These quantities are functions of both the stellar age at the earlier snapshot, $t_*$, and the time lag between snapshots, $\Delta t$. To analyse this, we place all stars in 2D bins of these quantities and compute the median\footnote{Note that since we are taking the modulus of the difference in \autoref{eq:DeltaJAbs} and then taking median here, this is similar to taking a square root of the mean of the square of the difference in actions at the two times. } in each bin; we use medians rather than means to ensure robustness against outliers. 
Before presenting the results, we note an important consequence of our definitions of $\Delta t$ and $t_*$. At small values of $t_* + \Delta t$, all stars formed during the simulation contribute to our analysis. However, only the stars formed near the start of the simulation contribute to the analysis at the limit $t_* + \Delta t < 464$ Myr (which is the time for which we run the simulations, excluding the initial transient phase that we discard). 

We show the binned changes in absolute action in \autoref{fig:heatmap}; in this plot, initial stellar age $t_*$ appears on the $x-$axis and the time lag $\Delta t$ on the $y-$axis. The expected behaviour is immediately visible -- small variations at short time intervals appear as the darkest colours in a horizontal band at the bottom of the plot. We see only a weak dependence on initial stellar age, as there are no very obvious vertical features in the heatmap.

\begin{figure}
        \includegraphics[width=0.47\textwidth]{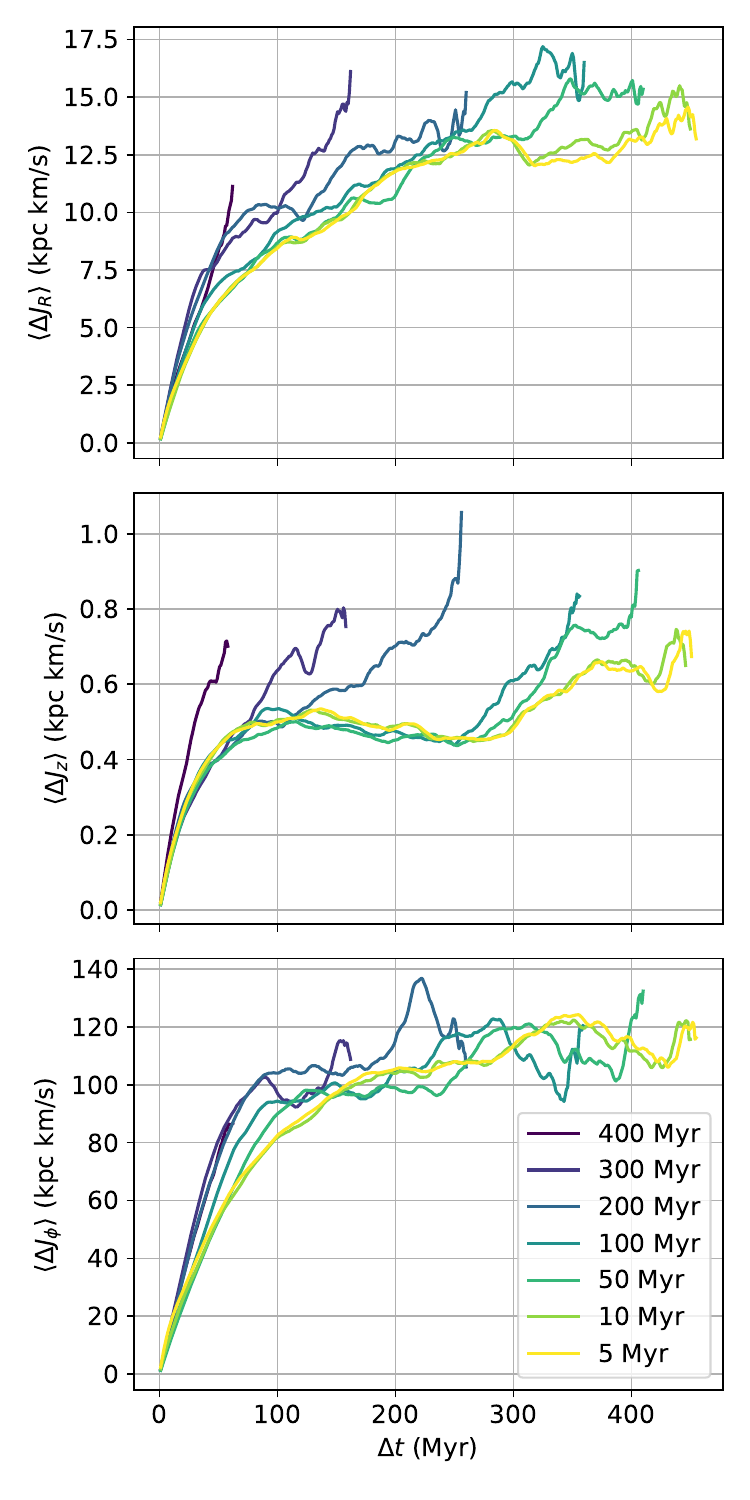}
        \caption{Median of the absolute change in stellar actions (row-wise in order: $\Delta J_R$, $\Delta J_z$, and $\Delta J_{\phi}$) for stars of different ages across the simulation as a function of time lag. Different initial stellar age groups are shown in different colours as indicated in the legend.}
        \label{fig:abs_J}
\end{figure}

To take a closer look, we extract vertical slices from the heatmap at $t_*= 5$, 10, 50, 100, 200, 300 and 400 Myr, as shown by the dashed lines in \autoref{fig:heatmap}. These slices, illustrated in \autoref{fig:abs_J}, show how the absolute change in actions depends on the time lag for stars of different initial ages. We see here that there is a weak dependence of $\Delta J$ on initial stellar age, but in a direction that is opposite what one might naively expect -- we expect younger stars to exhibit larger changes in their actions than older stars because they are more likely to reside near their natal molecular clouds \citep{lada&lada}, where ongoing interactions with dense gas or other stars can lead to greater dynamical perturbations. Conversely, older stars have typically migrated away from their birth regions and are less influenced by such perturbations. Contrary to these expectations, we instead find that older stars undergo more rapid changes to their actions, as indicated by the larger slopes in \autoref{fig:abs_J}. 

\begin{figure}
        \includegraphics[width=0.47\textwidth]{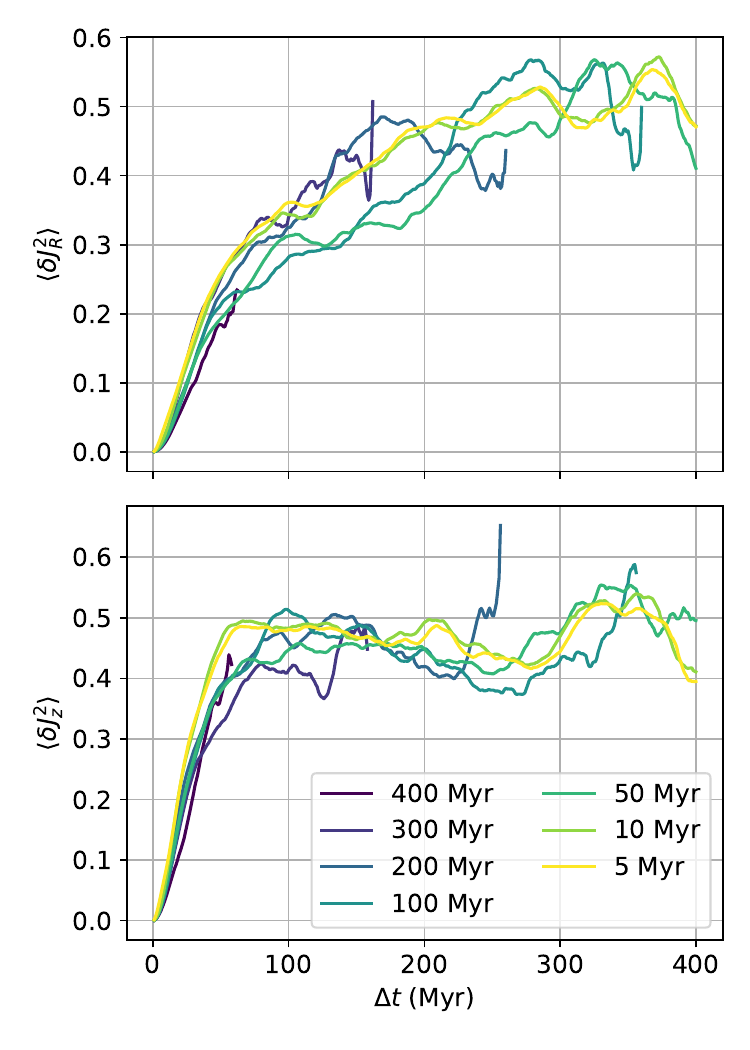}
        \caption{Same as the top two panels of \autoref{fig:abs_J}, but now showing the median square relative change in stellar actions (row-wise: $\delta J_R^2$ and $\delta J_z^2$).}
        \label{fig:rel_J}
\end{figure}

To understand why this occurs, we next examine relative rather than absolute changes in actions, focusing on $J_R$ and $J_z$ since $J_{\phi}$ is dominated by circular orbital motion and is less sensitive to dynamical perturbations. \autoref{fig:rel_J} shows the median square  relative changes of these two actions, $\delta J_R^2$ and $\delta J_z^2$, as a function of time lag for the same initial stellar ages as shown in \autoref{fig:abs_J}. In this plot, we see that the different initial stellar age bins collapse onto nearly a single line, suggesting a universal behaviour among stars of all ages.

The linear increase in $\delta J_R^2$ and $\delta J_Z^2$ with time before 100 Myr suggests that the process of stellar action change can be described as a random walk in the logarithm of the action, and that the process can therefore be approximated as diffusive. As discussed earlier, this diffusion is due to dynamical perturbations in the smooth, axisymmetric gravitational potential caused by structures such as giant molecular clouds and spiral arms, together with the large-scale acceleration of the disc due to galactic winds. 

\begin{figure}
        \includegraphics[width=0.47\textwidth]{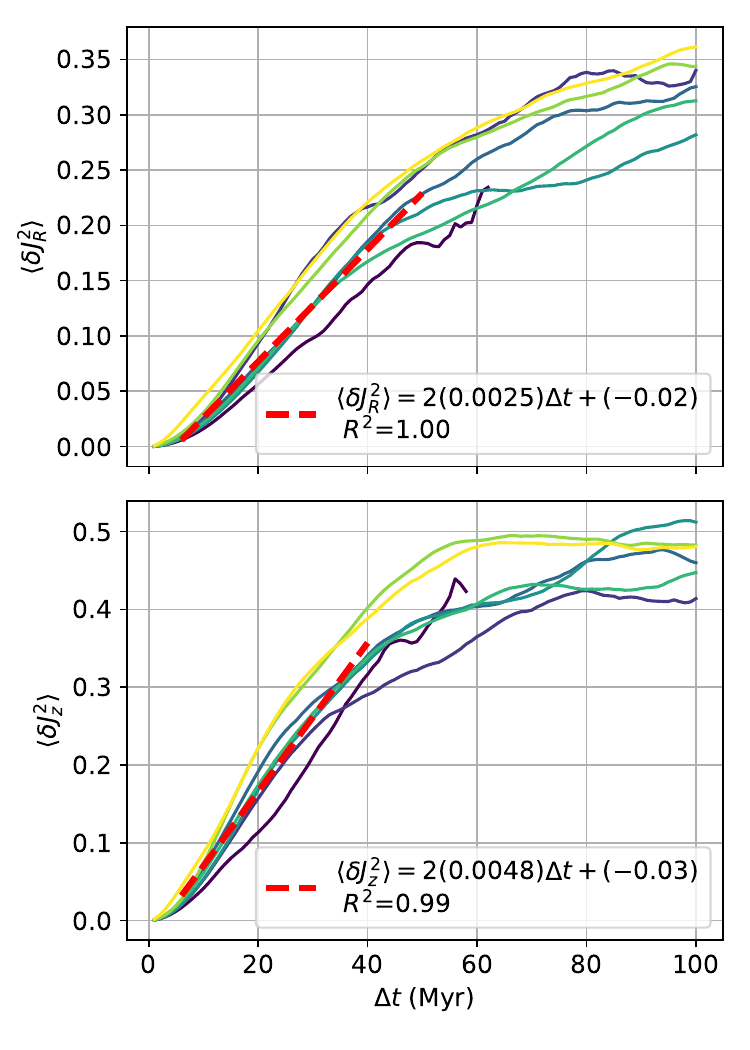}
        \caption{Same as \autoref{fig:rel_J}, but now zooming in on time lag $\Delta t$ $<100$ Myr. The red dashed lines indicate linear fits to the data, as described in the main text. The expressions and $R^2$ values for the fits are indicated in the legend, with times in units of Myr.
        }
        \label{fig:rel_J_fit}
\end{figure}

This clearly highlights a key result: actions are not conserved for individual stars, placing constraints on reconstructing stellar orbits using present-day actions. To quantify the rate of diffusion, we carry out a least-squares find of the data shown in \autoref{fig:rel_J} to a diffusion model of the form
\begin{equation}
    \label{eq:diffusion}
    \left \langle\delta J_R^2 \right\rangle = 2D_R \,\Delta t,
\end{equation}
and similarly for $z$, where here $D$ is the diffusion coefficient; we carry out this fit from $\Delta t = 5 - 50$ Myr for $D_R$ and for $5-40$ Myr for $D_z$, covering the age range when the behaviour is close to linear, and for the purpose of the fit we include all stars regardless of age $t_*$. We show our fits, together with the data, in \autoref{fig:rel_J_fit}. Our best-fitting values for the diffusion coefficients are $D_R=2.5$ Gyr$^{-1}$ and $D_z=4.8$ Gyr$^{-1}$. The inverses of these values correspond to characteristic diffusion timescales of approximately 400 Myr for $J_R$ and 200 Myr for $J_z$, meaning that over these timescales individual stars forget their individual actions. The shorter timescale for $J_z$ reflects stronger dynamical perturbations in the vertical direction. In both cases, however, we see that the curve in \autoref{fig:rel_J} flattens beyond $\approx 100$ Myr, indicating that actions have reached maximum decorrelation by this point, corresponding to around 50\% median square relative change. By this time, stellar actions are essentially drawn at random from the distribution of actions for stars of the appropriate age, and retain no memory of their values $\approx 100$ Myr earlier.

Note that the timescale that we measure here for changes in the actions of individual stars is much shorter (at least an order of magnitude) than the timescale for evolution of the population-median action or population velocity dispersion we find in \autoref{sec:comparison}. This difference arises because the two approaches measure fundamentally different processes. Population-level statistics such as AVRs or action distributions are sensitive only to changes across all stars, so for example if the actions of two stars were swapped these statistics would remain unchanged. By contrast, our absolute and relative action-change metrics defined in \autoref{eq:DeltaJAbs} and \autoref{eq:DeltaJrel}, even when summarised via their median over a large number of stars, would register a large change in such a swap. This explains how the population median action can require Gyr to evolve even while individual stars change their actions on a much smaller timescale as we see here. It also highlights that stability in the distribution of actions does not necessarily imply conservation of actions for individual stars.

While this is a good first-order description of the results, close examination of \autoref{fig:rel_J_fit} does reveal a weak age dependence of action diffusion: younger stars (represented by lighter colours) exhibit steeper slopes compared to older stars (darker colours) for both $J_R$ and $J_z$, as indicated by the fact that the lighter-coloured lines corresponding to the younger stars lie predominantly above the red dashed line representing the entire dataset, while the darker lines for older stars lie below it. This is consistent with our expected age dependence for action diffusion: younger stars are more susceptible to perturbations near their birth sites, changing their actions rapidly, while older stars' actions change comparatively slowly. However, we can now see that this is true only for \textit{relative actions}, while older stars experience faster changes in their \textit{absolute actions} because they have larger radial and vertical actions on average. The difference is not large, however.

\subsection{Dependence on local birth density}
\label{subsec:density}
\begin{figure}
        \includegraphics[width=0.47\textwidth]{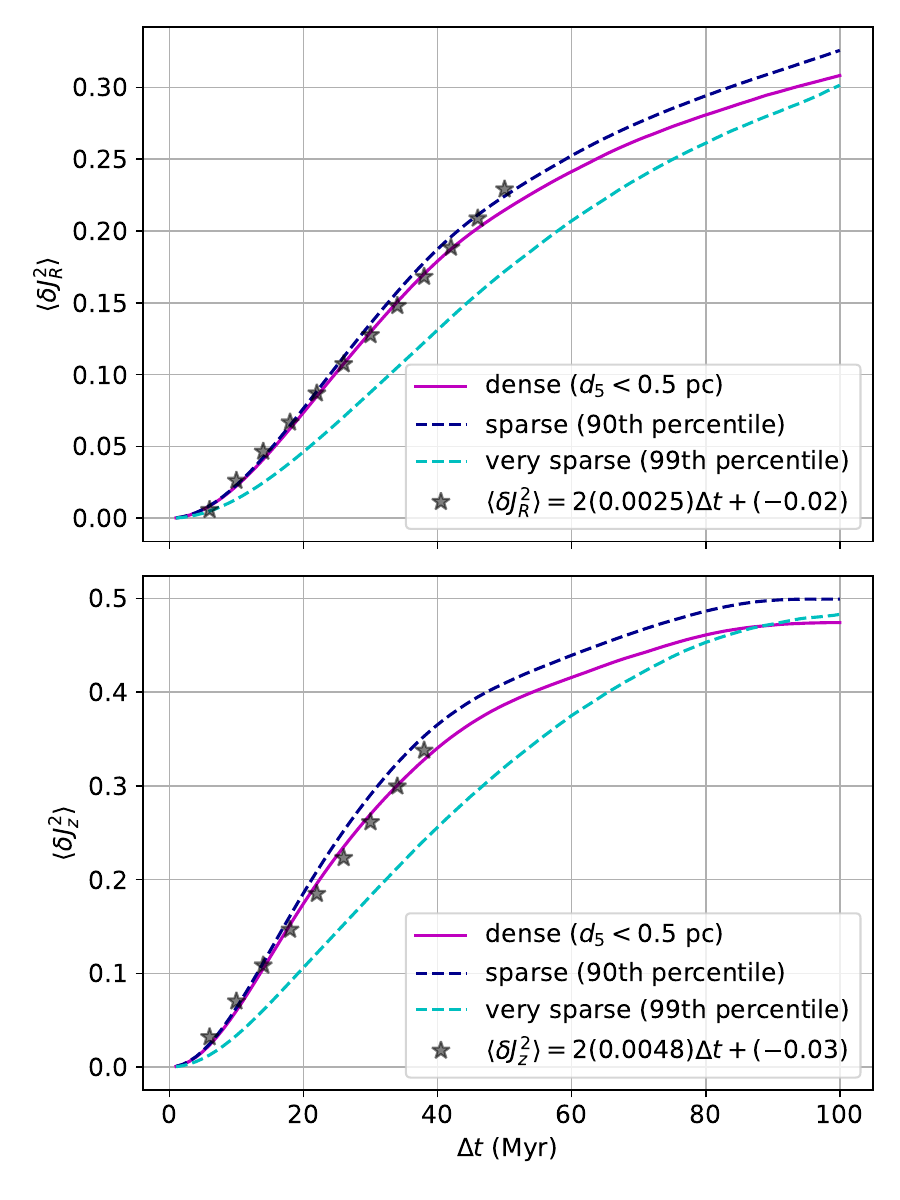}
        \caption{Square median  relative change in stellar actions (row-wise: $\delta J_R^2$ and $\delta J_z^2$) for stars born in regions of different density. The solid line shows stars born in the `dense' regions, while the dashed lines show those born in `sparse' (dark blue) `very sparse' (cyan) regions -- see main text for full definitions. The star markers indicate our linear fits to the full sample without subdividing by density, as shown in \autoref{fig:rel_J_fit}; the functional form of the fit is provided in the label, with $\Delta t$ in units of Myr.
        }
        \label{fig:density}
\end{figure}
The slightly more rapid diffusion of relative actions that we measure for younger stars suggests an environmental effect -- younger stars may diffuse more quickly because they are in closer proximity to denser regions with stronger local, non-axisymmetric gravitational pulls. However, tracking a star's density history over its entire lifetime or orbit on the timescales considered is complex, as any given star may transverse different regions of varying densities. Instead, to test whether density matters, we restrict our focus to the contribution of local density of the birth environment of the star to changes in its action. 

To achieve this, we define `newborn star cohorts' at each snapshot, consisting of all stars formed in the past 1 Myr. For each star in this cohort, we find the fifth nearest neighbour distance ($d_5$) to another star in the cohort as a proxy for the local density at which each star formed. Smaller $d_5$ values correspond to stars forming in denser environments where stars are packed more closely together. By combining these measurements across snapshots, we obtain a distribution of $d_5$ values, allowing us to assess the effect of environments in which stars are born. We find that the central parts of the distribution of $d_5$ are relatively smooth, with a 10th percentile value of 0.34 pc and 90th percentile value of 2.64 pc, thus spanning roughly a factor of 500 in stellar density (which scales as $d_5^{-3}$). There is also a low-density tail to the distribution, so that the 95th and 99th percentiles increase sharply to 87.2 pc and 1323.53 pc, respectively. Given that the smoothing length in our simulation is $\sim 0.5$ pc, we define `dense' birth regions as those where $d_5<0.5$ pc, moderately sparse regions as those for which $d_5$ is between 89th and 91st percentile value, and `very sparse' regions defined as those with $d_5$ beyond the 99th percentile, i.e., $d_5>1323.5$ pc. 

To see whether birth density affects the rate of action diffusion, we plot the square median relative change in radial and vertical actions, $\langle \delta J_R^2 \rangle$ and $\langle \delta J_z^2\rangle$, for stars in our three sample density bins in \autoref{fig:density}. The solid line represents stars that formed in dense environments, while the dashed lines correspond to those originating in sparse and very sparse regions -- dark blue for the 90th percentile in $d_5$ (2.64 pc) and cyan for the 99th percentile in $d_5$ (1323.5 pc). For reference, we also overlay the best-fit diffusion relation from the previous section (shown as star markers), which we compute for all stars regardless of their birth density. The results show that the stars formed in dense and sparse regions both follow the same trend as the overall dataset, closely matching the linear fit, and indicating no significant dependence on density. Stars born in very sparse regions exhibit smaller changes in action, particularly at lower $\Delta t$ values, consistent with the hypothesis that stars born in denser environments suffer more environmental perturbations and thus diffuse more rapidly. 
However, the effect is extremely weak, and affects only a small fraction of stars: there is no visible difference for stars at the 90th, or even 95th (not shown), percentile of density. Only the $\sim 1\%$ of stars born in the quietest, most isolated environments show significantly reduced rates of action diffusion. 
It is worth noting that while birth density does not appear to strongly influence action changes, this does not rule out the possibility that the density of the environment a star encounters throughout its orbit could play a role. A more comprehensive analysis would involve tracking the density of regions that stars pass through over time and correlating this with changes in their actions. However, such an investigation is beyond the scope of this paper.

\begin{figure*}
\centering
        \includegraphics[width=0.9\textwidth]{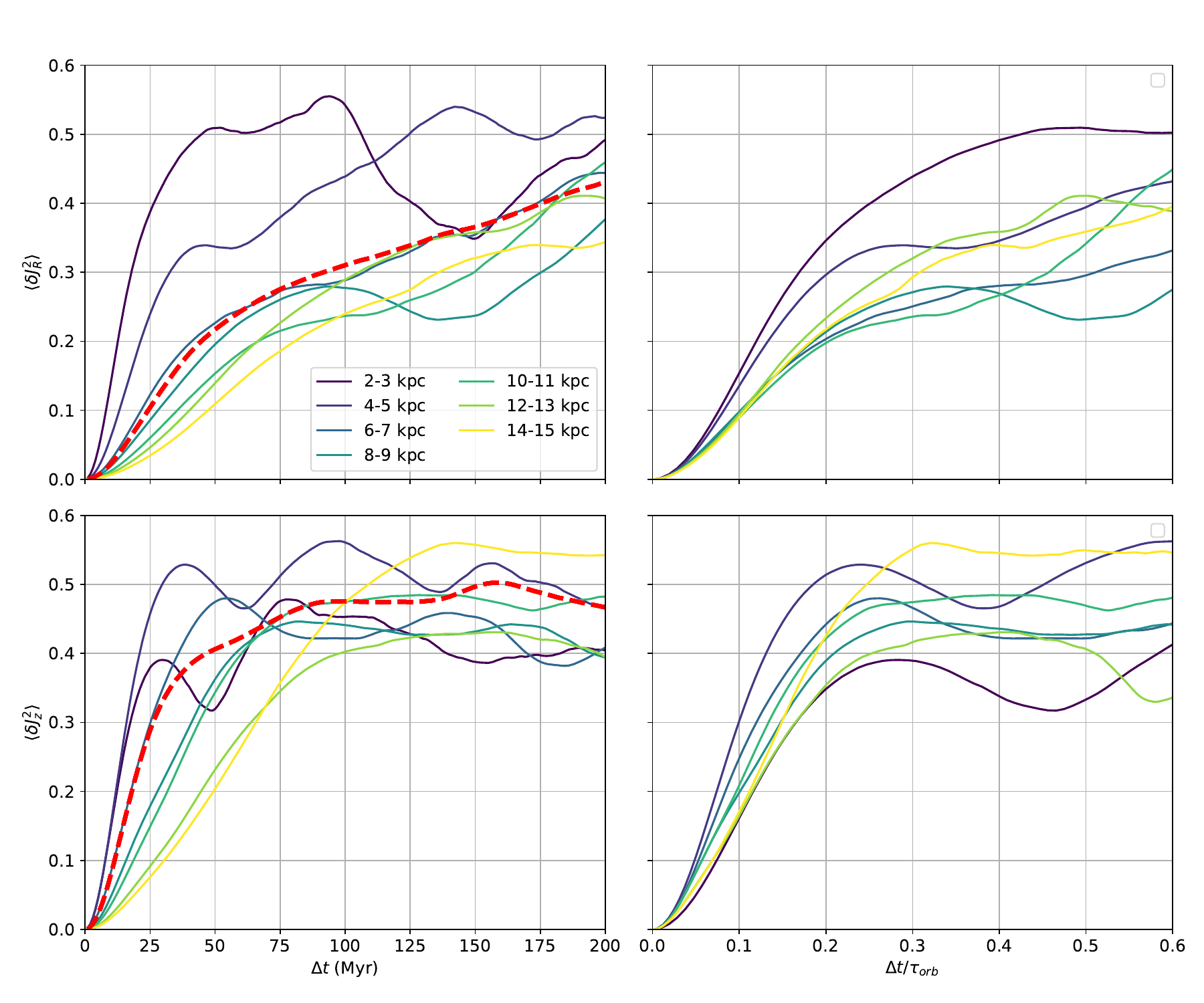}
        \caption{Square median relative change in stellar actions (row-wise: $\delta J_R^2$ and $\delta J_z^2$) for stars born in different radial bins. The birth radial bins are represented by different colours indicated in the legend. The red dashed line shows the values for the full sample without subdividing into radial bins. The left panels show the result in absolute time, while the right panels display the same data with time normalised by the orbital period ($\tau_{\text{orb}}$) in each radial bin.
        }
        \label{fig:radial_bins}
\end{figure*}

\subsection{Dependence on birth radius}
\label{subsec:radius}

Since disc structure and density, and hence frequency of perturbations, should vary with galactocentric radius, we next examine how rates of action diffusion vary with this quantity. We bin our stars by the galactocentric radius at which they are born, using 1 kpc-wide bins, and show the square of the median relative change in radial and vertical actions, $\langle \delta_R^2 \rangle$ and $\langle\delta J_z^2\rangle$, for different radial bins in \autoref{fig:radial_bins}. The left panels present the result in absolute time, while the right panels show the same with the time normalised by the orbital period at the centre of the radial bin.

In absolute time, the rate of action change varies significantly with radius. The outer disc regions consistently show slower evolution compared to the inner disc for both $R$ and $z$ actions. This would suggest that the processes driving action diffusion are less efficient at larger radii. However, we must keep in mind that the orbital period is higher at higher radii. When time is normalised by the orbital period ($\tau_{\text{orb}}$), this trend becomes much less pronounced as seen in the right panels of \autoref{fig:radial_bins}, suggesting that to a first-order approximation, the rate of action diffusion is constant when time is measured in units of the orbital period. However, a residual difference still remains for cohorts of stars born in different radial bins even after dividing by the orbital period, indicating that the gravitational perturbations responsible for action evolution are less frequent in the outer disc than in the inner regions even accounting for the longer orbital period. This is not surprising given that the stellar density is higher and gravitational interactions with GMCs and spiral arms are more frequent in the inner disc regions. 
Additionally, it is worth noting that within 60\% of the orbital period, squares of relative change in actions are already between $30-50\%$. This clearly highlights the limitation of using methods that rely on the long-term conservation of action for reconstructing past orbits or predicting future trajectories. Hence, for practical applications of actions, careful consideration of the timescales over which they are assumed to be invariant is required.

\subsection{Comparison with initial stars}
\label{subsec:initial_stars_comparison}

\begin{figure}
        \includegraphics[width=0.47\textwidth]{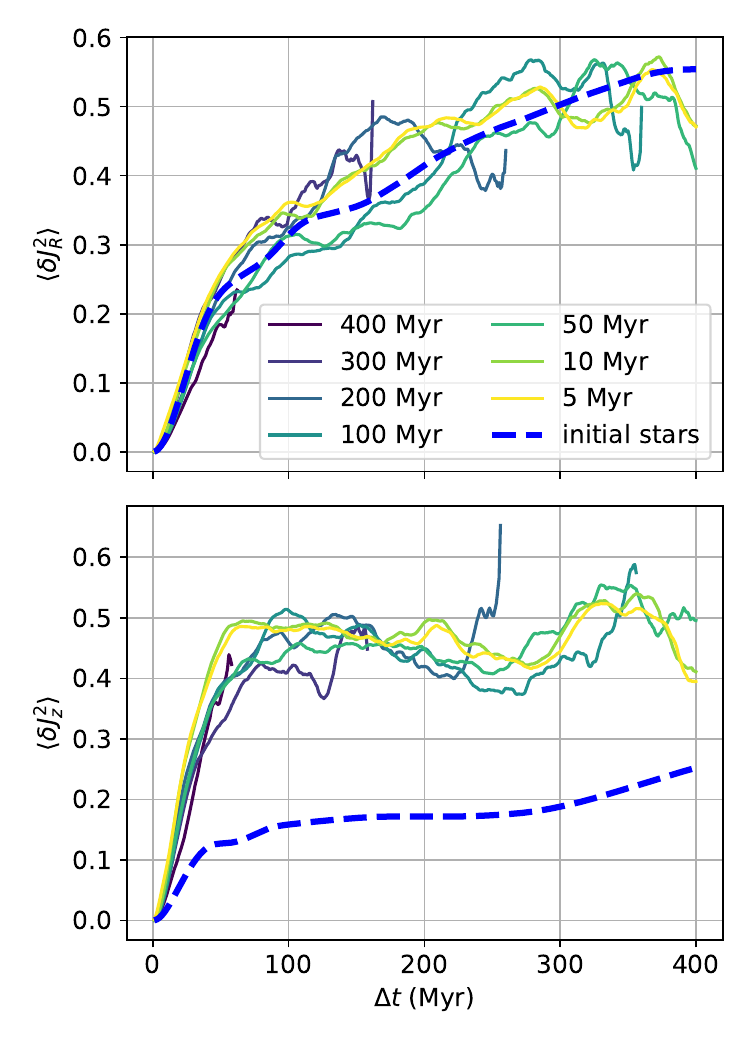}
        \caption{Same as \autoref{fig:rel_J}, overlaid with a blue dashed line that represents median square relative change in stellar actions for the initial stars (row-wise: $\delta J_R^2$ and $\delta J_z^2$)}
        \label{fig:init_J}
\end{figure}

As described in \autoref{subsec:initial_cond}, the \citetalias{wibking2023} simulations include a population of stellar disc particles that are already present in the initial conditions of \citetalias{chuhan2024}. We refer to these pre-existing particles as `initial stars'. In contrast, the star particles discussed so far are those born during the simulation and hence are all younger than 464 Myr. To understand the underlying mechanisms driving action evolution, it is illuminating to compare these two populations. Since the potential grid used to calculate action (see \autoref{sec:method} and \autoref{app:action_calc}) extends only to the disc, this comparison is restricted to stars within the disc region.

The key differences between these two populations are their vertical height distributions and their distribution relative to dense gas. Initial stars have positions that are uncorrelated with gas, while stars that form in the simulation are necessarily in close proximity to dense gas structures at least at birth. Similarly, the root mean square height $\langle z^2\rangle^{1/2}$ of initial stars remains between $380-405$ pc during the simulation, whereas, as already mentioned, for stars born during the simulation, it is much lower, between $130-200$ pc. Thus the initial stars spend much more time away from the midplane while their younger counterparts remain confined to the midplane. 

In \autoref{fig:init_J}, we show the median square relative changes in radial and vertical actions, $\delta J_R^2$ and $\delta J_Z^2$, as a function of time lag for stars of different initial stellar ages born in the simulation (same as in \autoref{fig:rel_J}) along with the same for initial stars, represented by the blue dashed line. Despite the differences between the two populations, the change in radial action over time is quite similar. The fact that even the oldest stars experience a comparable level of radial action change to newly born stars suggests that this change is driven by a mechanism that is present throughout the disc and does not depend strongly on birth conditions and proximity to small-scale gas structures. The natural candidate that meets this requirements are transient spiral arms, which are known to be a key feature of disc dynamics and which extend over a range of vertical heights, since for a Milky Way-like disc they involve strongly-coupled perturbations in both gas and stars \citep[e.g.,][]{Romeo17a}.

On the other hand, the evolution of the vertical action differs significantly between the two populations. For initial stars, the vertical action remains nearly conserved, flattening out to about $20\%$ change by the end of the simulation. In contrast, as seen already in previous sections, the stars born during the simulation show much larger changes in the vertical action. This suggests that the mechanism for out-of-plane action evolution is different from the mechanism for in-plane action evolution, and relies on proximity to gaseous structures in the midplane. This leaves  scattering by GMCs, which are concentrated near the midplane, as the natural candidate: initial stars, which spend most of their time at greater heights, experience fewer encounters with GMCs and thus retain more of their initial vertical actions.

\begin{figure}
        \includegraphics[width=0.47\textwidth]{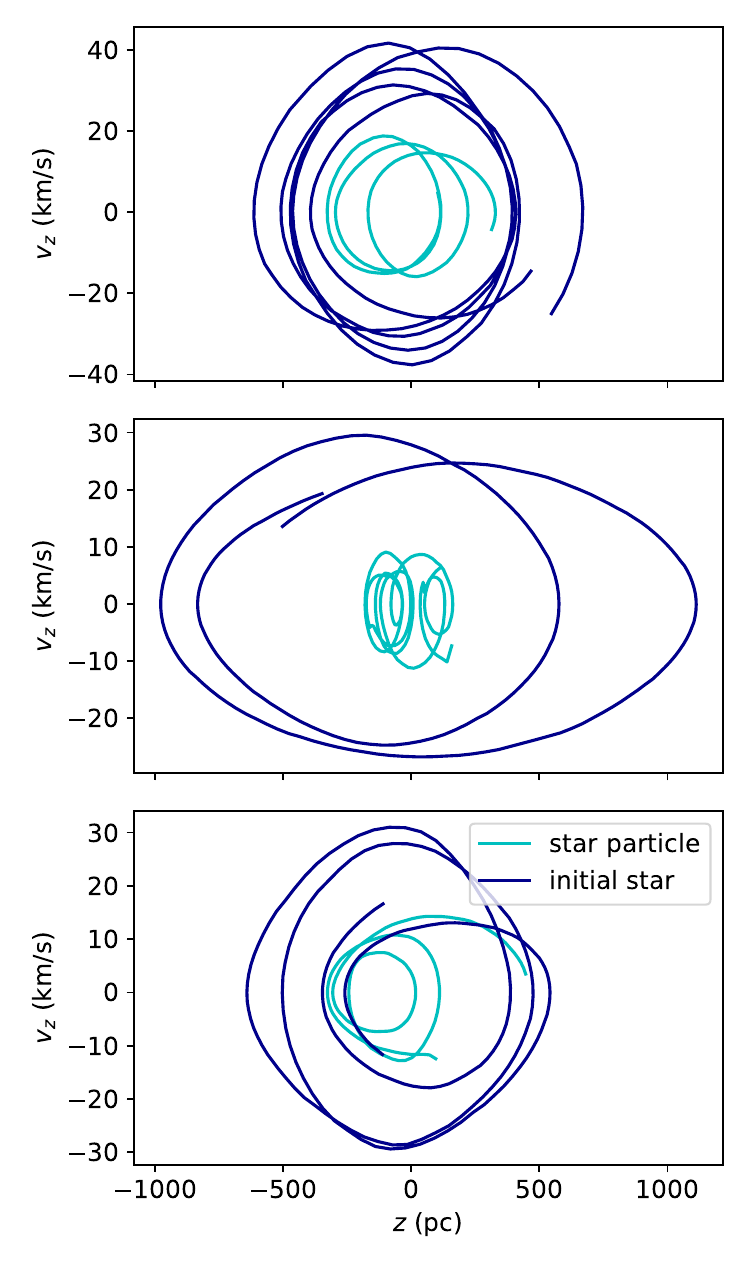}
        \caption{Examples of phase-space orbits for initial stars (cyan) and star particles born in the simulation (dark blue). The x-axis shows their vertical positions in parsecs, while the y-axis shows vertical velocities in km/s.}
        \label{fig:zvsvz}
\end{figure}

To further confirm our hypotheses, we examine phase-space orbits of sample initial stars and stars born in the simulation in \autoref{fig:zvsvz}. We select for this purpose stars whose action evolution lies close to the median values of their respective populations, ensuring that these orbits are typical trajectories for both the categories. Since we applied a low-pass filter to the action time series (see \autoref{subsec: action evolution}), we handpick stars that do not seem to be part of clusters to minimise the impact of filtering. The figure shows that orbits do confirm the expected trend: initial stars with larger actions have nearly-closed orbits, as expected for a conserved action, while the stars born during the simulation follow more irregular trajectories, which is consistent with stronger vertical action evolution for the younger population.

\section{Conclusion}
\label{sec:conclusion}
In this paper, we use high-resolution MHD simulations to study the evolution of actions in young stars from Myr to Gyr timescales while self-consistently incorporating gas dynamics, star formation, and a live stellar disc. Our key findings can be summarised as follows:

\begin{itemize}

    \item The distributions of actions over our simulation time of $\approx 500$ Myr remains almost constant, broadening only slightly. For parts of the disc interior to $\approx 12$ kpc, and thus interior to the warp in the Milky Way disc, we find good agreement between the rate of broadening in the simulation  and that observed in the Milky Way. We also find that the age-velocity relation in our simulations is relatively flat, with only modest increases in the stellar velocity dispersion over the simulation time.
    \item Individual stellar actions, however, are not conserved and instead undergo relatively rapid evolution. The effect is strong enough that over timescales of a few hundred Myr stellar actions become fully decorrelated, and it is no longer possible to trace stars backward under the assumption of an axisymmetric time-steady potential. Contrary to expectation, absolute actions change more rapidly for older stars than for younger ones. However, we show that if we instead consider actions normalised by their initial values, younger stars indeed experience more rapid evolution as expected from the stronger perturbations near their birth environments.
    \item At ages $\lesssim 100$ Myr before stellar actions are fully decorrelated, the square median relative change in action increases roughly linearly with time lag, suggesting a random walk in logarithmic action space. We fit a diffusion model to these trends and find diffusion coefficients of $D_R = 2.5$ Gyr$^{-1}$ and $D_z = 4.5$ Gyr$^{-1}$ for radial and vertical actions respectively. 
    \item We find that for the great majority of stars, the density of the stellar birth environment has little effect on the rate of action diffusion. However, this rate does vary with galactocentric radius. To first order, the diffusion coefficient simply scales as the inverse of the local galactic rotation period, and to second order, action diffusion is somewhat slower at larger radii even normalised to the orbital period. This is expected, given the lower stellar density and reduced frequency of encounters with GMCs and spiral arms in the outer disc.
    \item Comparing newborn stars to an older population with a factor of $\approx 2-3$ larger scale height, we find both populations have similar rates of evolution for radial action evolution, but that the vertical action evolves significantly more slowly for the older, more vertically-extended population. This suggests that different mechanisms drive changes in radial and vertical actions: predominantly transient spiral arms for the radial action, and predominantly scattering of the thin gas disc for the vertical action.
\end{itemize}

While the rapid action evolution we measure might at first seem surprising, it is important to remember that many previous studies of stellar action conservation have been conducted in the context of a purely collisionless stellar disc, and even in these cases actions are not perfectly conserved \citep[e.g.,][]{solway2012,mikkola2020}. Other studies that include gas find that its gravitational influence strongly scatters young stars \citep[e.g.,][]{fujimoto2023}, and the general importance of gas becomes apparent if we recall that, while gas makes up only $\sim 15\%$ of the mass of the Milky Way disc, its smaller scale height means that at the midplane gas and stars contribute about equally to the vertical gravitational acceleration \citep[e.g.,][]{Mckee15a, Krumholz18b}. In this context it is worth considering as an example the Local Bubble, which is effectively a void in the local ISM (interstellar medium), which otherwise has a mean density of $\sim 1$ atom cm$^{-3}$, several hundred pc across \citep{linsky2022, zucker2022b, localbubble2024}. The free-fall time of this structure is (\autoref{eq:ff_time}) $\approx 51.5$ Myr,  comparable to the epicyclic periods in the Solar neighbourhood \citep[$T_{\kappa}\approx170$ Myr and $T_{\nu}\approx85$ Myr;][]{binney2008}, and its size means stars passing through this region experience a non-axisymmetric gravitational perturbations over a significant portion of their orbit. It is therefore not surprising that a structure of this size should be able to perturb stellar orbits significantly.

Our results imply that disc stars, unlike halo stars, do not follow regular, predictable orbits, and are subject to more rapid and erratic action evolution. Even in the halo, orbit reconstruction is difficult: \citet{arora2022} find that even in the absence of major mergers, integration of orbits in a time-independent potential renders orbit reconstruction of stellar streams unreliable after only 0.5--1 Gyr. This inability of orbit recovery is only likely to be more severe in the disc, where dynamical times are shorter and perturbations from gas and substructure are more frequent and intense. This idea is further supported by \citet{kamdar2021}, who study stellar streams in a dynamical model of Milky Way-like disc and find that even with perfect information about the non-axisymmetric structures in the galaxy (spiral arms and a bar), the fraction of stream stars that could be reliably integrated backward in time drops below half within 100 Myr. With integration in only axisymmetric potential, the fraction drops below half within few tens of Myr. Notably, their work uses an analytical potential with idealised structure, whereas our simulation includes live ISM, self-consistent star formation and MHD physics. Therefore, in the more realistic galactic environment of our simulation, we conclude that reliable stellar orbit reconstruction relative to the disc  past $\sim 100$ Myr is not feasible.

However, this does not necessarily mean that star clusters are entirely irrecoverable. While individual stars undergo a random walk in logarithmic action space, our preliminary analysis indicates that stars born together tend to remain correlated with \textit{each other} in action space over time. This correlation is not surprising, given that much of the change in action appears to be driven by relatively large-scale perturbations such as the Local Bubble -- stars born closer to one another than the characteristic sizes of such structures are subjected to similar perturbations, and therefore should experience comparable changes in their actions and remain together in action space. In fact, the errors introduced by non-conservation of actions are likely to be similar for stars that form close together, meaning that while their individual orbits may be inaccurate, backward integration will still place them near the same location -- enabling recovery of the cluster, though not its orbit relative to the Galactic disc. In essence, while stars drift randomly in action space, those that form together ``hold hands'' as they move, thus preserving clustering signatures in action space that could still be used to reconstruct star clusters and associations, if not to trace them back to their formation locations. We will explore this idea further in Paper II of this series.

\section*{Acknowledgements}
We thank the anonymous referee for their comments that improved the clarity of this work. 
This research was undertaken with the assistance of resources from the National Computational Infrastructure (NCI Australia) and the Pawsey Supercomputing Centre, NCRIS-enabled capabilities supported by the Australian Government, through award jh2. AA and MRK acknowledge support from the Australian Research Council through Laureate Fellowship FL220100020. MI wishes to thank HW Rix for lunchtime discussions in 2019 that motivated some of this work, and the Australia-Germany Joint Research Co-operation Scheme of Universities Australia for funding this travel.


\section*{Data Availability}
\label{sec:data}
The codes and a subset of selected intermediate data used in this study are available at \url{https://github.com/aruuniima/stellar-actions-I} and \url{https://www.mso.anu.edu.au/~arunima/stellar-actions-I-data/}. The initial simulation outputs and potential fits are not included in the public repository due to their size, but will be provided by the authors upon reasonable request.



\bibliographystyle{mnras}
\bibliography{references} 



\appendix

\section{Comparison of epicyclic actions with more accurate action estimates}
\label{app:galpy}

To assess the accuracy of the epicyclic approximation used throughout this work, we select a random sample of 10,000 stars from the final snapshot of our simulations and compare the actions we compute for these stars using the epicyclic approximation to those we obtain using a much more accurate (but considerably more computationally-intensive) action estimate based on the St\"ackel approximation as implemented in the \texttt{galpy} package \citep{bovy2015}. This method has been widely adopted in both observational and simulation-based studies of Galactic dynamics \citep[e.g.,][]{frankel2018, trick2019, jia2023,palicio2025}.

Our first step in this comparison is to construct the axisymmetric potential.
Most previous studies have relied on the time-independent analytic Milky Way-like potentials provided by \texttt{galpy}. However, we have seen that our simulation, with its live stellar and gas disk, does not have an entirely constant potential. To remain consistent with the simulations, therefore, we instead extract the gravitational potential and acceleration fields directly from our simulation snapshot and input them into \texttt{galpy} as a custom potential. We do so as follows: first,  we extract the positions, masses, and smoothing lengths of all particles -- stars, gas, and dark matter -- and  use \texttt{pytreegrav} \citep{pytreegrav2021} to compute the resulting gravitational potential and acceleration on a regular cylindrical grid in $(R, z, \phi)$ with dimensions $(40000, 6000, 16)$, covering the range $R \in [0.1\ \text{pc}, 40\ \text{kpc}]$, $z \in [-3, 3]\ \text{kpc}$, and $\phi \in [0, 2\pi]$. To ensure axisymmetry, we average the potential and both the radial and vertical accelerations over $\phi$, resulting in a 2D $(R, z)$ grid. As in our epicyclic approximation pipeline, we next smooth these grids to facilitate accurate numerical differentiation using the same smoothing approach as described in \aref{app:action_calc} for the potential used in the epicyclic approximation -- see accompanying code linked in Data Availability for the full method. We then construct 2D spline fits for the potential, radial force, vertical force, and the $R$- and $z$-derivatives of the force, which are required for the numerical intgegrated used in \texttt{galpy}'s \texttt{actionAngleStaeckel} routine.

For each of our 10,000 sample stars, following \texttt{galpy} documentation, we integrate each star’s orbit in this potential for 50 internal time units (corresponding to $\approx 1.9$ Gyr) and use the resulting orbit data to estimate the local focal parameter $\Delta$ using the \texttt{estimateDeltaStaeckel} function. With the potential and $\Delta$ in hand, we then calculate actions using \texttt{actionAngleStaeckel}. This method successfully returns actions for $\approx 70 \%$ of the sample.\footnote{We have verified that the success rate is consistently around this value by repeating the process multiple times with different sample stars.} The failures are due to \texttt{galpy}'s inability to estimate $\Delta$, likely stemming from numerical noise in the derivatives. However, since the action computation itself is generally robust to such errors, for these failed cases we simply fix $\Delta = 0.5$, a typical value for disc orbits. With this substitution, the success rate increases to 98\%.

\begin{figure}
        \includegraphics[width=0.47\textwidth]{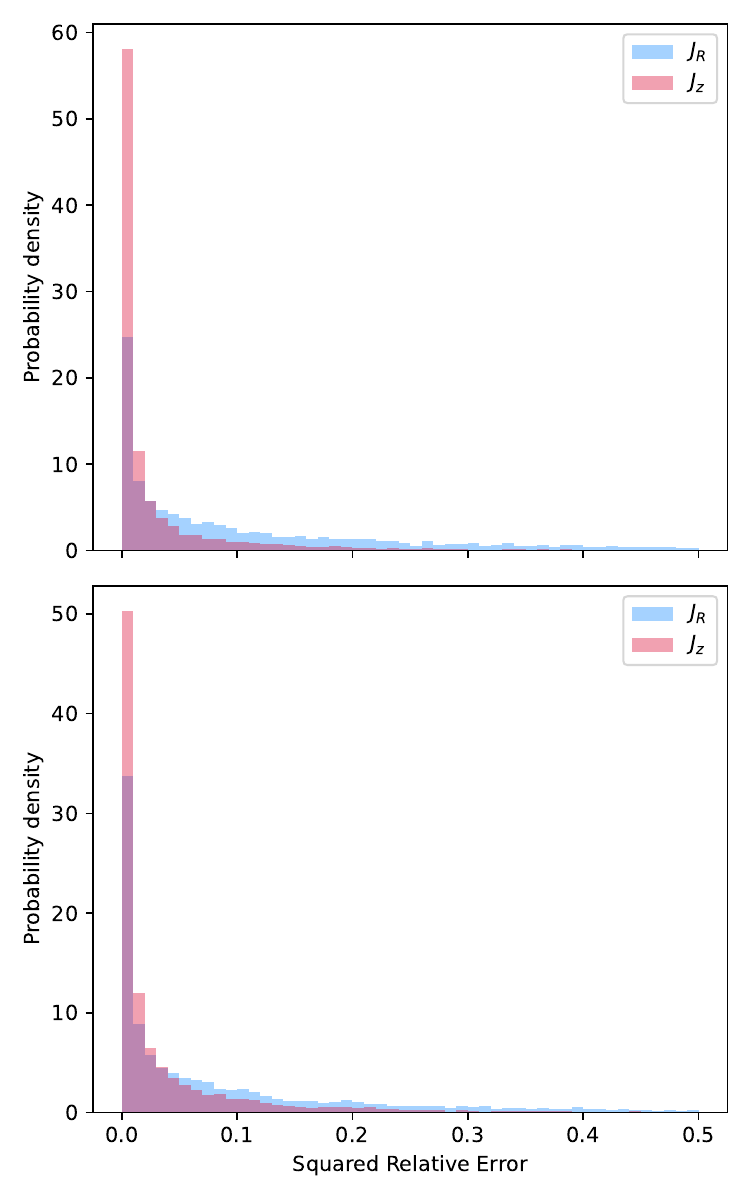}
        \caption{The probability distribution of the squared relative error of epicyclic actions with respect to the more accurately determined actions for the 75\% success rate sample (upper row) and the 98\% success rate sample (lower row). The radial and vertical action errors are shown in blue and pink colours, respectively.}
        \label{fig:action_compare}
\end{figure}

We compare the epicyclic and Stäckel-based action estimates in \autoref{fig:action_compare}, showing both the $\approx 70\%$ of the sample for which \texttt{galpy} works ``out of the box'' and the expanded sample using fixed $\Delta$. We find that the median squared relative errors in $J_R$ and $J_z$ for the 70\% sample are 0.08 and 0.01, respectively, and 0.05 and 0.01, respectively, for the expanded sample. Errors in $J_\phi$ are negligible ($<4 \times 10^{-14}$ kpc km/s for all stars) and are not shown. These figures are consistent with the small differences we expect based on the findings of \citet{solway2012}, and are much smaller than the variations in mean-square action that we find in the main text. This confirms that the epicyclic approximation is sufficient for our purposes.

\section{Computing epicyclic frequencies}
\label{app:action_calc}
In this appendix, we describe in detail the procedure that we use for calculating epicyclic frequencies $\kappa$ and $\nu$ for each star particle. We begin with the 2D potential grid that spans from 0.1 pc to 20 kpc in the radial direction and from $-1$ to $+1$ kpc in the vertical direction, computed as described in \autoref{subsec: grav pot profile}, and we must evaluate partial derivatives of this potential to compute $\kappa$ and $\nu$. Naive numerical evaluation of these derivatives yields extremely noisy estimates, and we must therefore adopt a smoothing procedure.

The guiding radius $R_g$ and radial epicyclic frequency depend on $\partial\Phi/\partial R$ and $\partial^2 \Phi /\partial R^2$ evaluated at $z=0$. Our first step to estimate these quantities is to extract the potential grid values within $\pm 2$ pc of the midplane, corresponding to five grid points in the vertical direction, and average them to produce a one-dimensional function of $R$. This averaging reduces the noise. We next apply a Gaussian filter\footnote{This and all subsequent operations use the implementations provided in SciPy version 1.11.4; see \citet{SciPy20a}.} with a standard deviation of 30 pc using \texttt{`nearest'} mode to prevent artificial edge effects; this further smooths the data. Finally, we fit the smoothed data with a B-spline with smoothing parameter $s=200$ to obtain a continuous, differentiable function $\Phi (R, z=0)$. 

With this smooth function at hand, we are now prepared to evaluate the radial derivatives of $\Phi$ and the quantities that depend upon them. We first define a new radial grid with a resolution of 0.5 pc with its inner edge placed at the larger of 2 kpc and 5 kpc inward from the smallest radial position of any star at the time we are sampling, and its outer edge at the smaller of 17.5 kpc and 5 kpc outward from the largest stellar radial position; this ensures that we have good coverage over the full disc, but that we are not extrapolating to radii where we sample the potential poorly. We then evaluate $\Omega_i$ at each radial grid point $R_i$ from \autoref{eq:Omega} using our B-spline representation of $\Phi(R)$. The corresponding specific angular momentum at each grid point is $L_{z,i} = R_i^2 \Omega_i$, and thus our grid represents a set of $(R_i, L_{z,i})$ pairs. Since the guiding radius $R_g$ for each star is defined implicitly by its angular momentum $L_{z,*}$ and the condition $L_{z,*} = R_g^2 \Omega(R_g)$, we can use the $(R_i, L_{z,i})$ pairs as a lookup table to find the guiding radius corresponding to any $R_g$. In practice we do this by defining a new B-spline function $R_g(L_z)$ from our $(R_i, L_{z,i})$ pairs using a smoothing factor $s=2$, and use the resulting function to evaluate $R_g$ for every star. We then drop from our sample stars for which $R_g$ lies outside the range $2-16$ kpc, on the grounds that they lie too close to our grid edge to be reliable.

Once the guiding radii are determined, we use them to compute the radial epicyclic frequencies via an analogous strategy. We calculate $\kappa (R_g)$ using \autoref{eq:kappa} at each grid point, again using our smooth B-spline approximation to $\Phi(R,z=0)$ to evaluate derivatives. This yields a set of $(R_i, \kappa_i)$ pairs, to which we apply another B-spline fit with smoothing factor $s=3$ to obtain a smooth function $\kappa(R)$. We plug the guiding radii for each star in our sample into this function to obtain $\kappa$ values for each of them. 

The calculation of the vertical epicyclic frequency $\nu$ follows a similar but slightly more complex process, since we require the second derivative of $\Phi (z)$ at each star's guiding radius (see \autoref{eq:nu}), rather than for the radial case which only requires derivatives of $\Phi (R)$ evaluated at $z=0$. To achieve this, we construct a sparse radial grid with a resolution of 200 pc, spanning the range $2-16$ kpc to match the guiding radius constraints that we imposed earlier. At each radial grid point, we extract the potential from the 2D grid by averaging over all radial values within $\pm 1$ pc of the sparse grid point. This averaging, as in the radial case, reduces local noise and provides a better estimate of the galactic potential as a function of $z$. We then smooth the resulting $\Phi(z)$ data by applying a Savitzky-Golay filter with a window length of 101 pc and a polynomial order of 2, which smooths the data while preserving the underlying curvature, followed by a Gaussian filter with standard deviation of 2 pc, which we apply three times. Finally, we use the smoothed data as input to a B-spline fit with $s=50$ to generate a smooth, continuously-differentiable functional form of $\Phi(z)$ at each sparse radial grid point. We use this smooth function to evaluate $\partial^2 \Phi/\partial z^2$ at $z=0$, the quantity required to evaluate $\nu$ using \autoref{eq:nu}. This yields a set of $(R_i, \nu_i)$ pairs on our sparse radial grid. We fit these data with a B-spline fit with $s=1000$ to obtain a continuous function $\nu(R)$. Finally, we determine $\nu$ for each star by inputting its guiding radius into this function.


\bsp	
\label{lastpage}
\end{document}